\documentclass[aps,showpacs,twocolumn,longbibliography,superscriptaddress]{revtex4-1}
\pdfoutput=1
\usepackage[utf8]{inputenc}
\usepackage[english]{babel}
\usepackage[T1]{fontenc}
\usepackage{amsmath}
\usepackage{xcolor}
\colorlet{myPurple}{blue!40!red}
\colorlet{myCyan}{cyan!60!gray}
\colorlet{myRed}{blue!55!gray}
\usepackage{tikz}
\usepackage{pgfplots}
\pgfplotsset{compat=1.14}
\usepackage[colorlinks=true,citecolor=myRed,urlcolor=myRed,linkcolor=myRed]{hyperref}
\usepackage{exscale}
\usepackage{bbm}
\usepackage{graphicx}
\usepackage{amsmath}
\usepackage{latexsym}
\usepackage{amsfonts}
\usepackage{amssymb}
\usepackage{times}
\usepackage[T1]{fontenc}
\usepackage{amsthm}
\usepackage{enumerate}
\usepackage{bbold}
\usepackage{color}
\usepackage{nicefrac}
\usepackage{changes}

\newcommand{\sket}[1]{{\ensuremath{\lvert#1\rangle}}}
\newcommand{\lket}[1]{{\ensuremath{\left\lvert#1\right\rangle}}}
\newcommand{\ket}[1]{\if@display\lket{#1}\else\sket{#1}\fi}

\newcommand{\sbra}[1]{{\ensuremath{\langle#1\rvert}}}
\newcommand{\lbra}[1]{{\ensuremath{\left\langle#1\right\rvert}}}
\newcommand{\bra}[1]{\if@display\lbra{#1}\else\sbra{#1}\fi}

\newcommand{\sbraket}[2]{{\ensuremath{\langle#1\rvert#2\rangle}}}
\newcommand{\lbraket}[2]{{\ensuremath{\left\langle#1\!\left\rvert\vphantom{#1}#2\right.\!\right\rangle}}}
\newcommand{\braket}[2]{\if@display\lbraket{#1}{#2}\else\sbraket{#1}{#2}\fi}

\newcommand{\sketbra}[2]{{\ensuremath{\lvert #1\rangle\!\langle #2\rvert}}}
\newcommand{\lketbra}[2]{{\ensuremath{\left\lvert #1\right\rangle\!\!\left\langle #2\right\rvert}}}
\newcommand{\ketbra}[2]{\if@display\lketbra{#1}{#2}\else\sketbra{#1}{#2}\fi}

\newcommand{\tr}{\textrm{Tr}}
\newcommand{\da}{\dagger}

\newcommand{\rA}{\text{A}}
\newcommand{\rB}{\text{B}}

\newcommand{\rX}{\text{X}}
\newcommand{\M}{\mathsf{M}}
\newcommand{\N}{\mathsf{N}}

\usepackage{tikz}
\usepackage{lipsum}
\theoremstyle{plain}
\newtheorem{thm}{Theorem}

\newtheorem{cor}[thm]{Corollary}

\usepackage{graphicx}
\usepackage{bm}
\usepackage{dsfont}
\usepackage{tikz}
\usepackage[T1]{fontenc}
\usepackage{amsthm}
\usepackage{array}
\usepackage{amssymb}
\usepackage{amsfonts}
\usepackage{cancel}
\usepackage[toc,page]{appendix}
\usepackage{multirow}
\usepackage{color}
\usepackage{calrsfs}
\usetikzlibrary{backgrounds,decorations.pathreplacing,calc}

\begin{document}

\title{Self-testing and certification using trusted quantum inputs}

\author{Ivan \v{S}upi\'{c}}
\email{ivan.supic@unige.ch}
\affiliation{Département de Physique Appliquée, Université de Genève, 1211 Genève, Switzerland}
\author{Matty J. Hoban}
\affiliation{Department of Computing, Goldsmiths, University of London, New Cross, London SE14 6NW, United Kingdom}
\author{Laia Domingo Colomer}
\affiliation{ Universitat Aut\'{o}noma de Barcelona,  ES-08193 Bellaterra (Barcelona), Spain}
\author{Antonio Ac\'{i}n}
\affiliation{ICFO-Institut de Ciencies Fotoniques, The Barcelona Institute of Science and Technology, 08860 Castelldefels (Barcelona), Spain}
\affiliation{ICREA - Instituci\'{o} Catalana de Recerca i Estudis Avancats, 08011 Barcelona, Spain}

\date{\today}

\begin{abstract}
Device-independent certification of quantum devices is of crucial importance for the development of secure quantum information protocols. So far, the most studied scenario corresponds to a system consisting of different non-characterized devices that observers probe with classical inputs to obtain classical outputs. The certification of relevant quantum properties follows from the observation of correlations between these events that do not have a classical counterpart. In the fully device-independent scenario no assumptions are made on the devices and therefore their non-classicality follows from Bell non-locality. There exist other scenarios, known as semi-device-independent, in which assumptions are made on the devices, such as their dimension, and non-classicality is associated to the observation of other types of correlations with no classical analogue. More recently, the use of trusted quantum inputs for certification has been introduced. The goal of this work is to study the power of this formalism and describe self-testing protocols in various settings using trusted quantum inputs. We also relate these different types of self-testing to some of the most basic quantum information protocols, such as quantum teleportation. Finally, we apply our findings to quantum networks and provide methods for estimating the quality of the whole network, as well as of parts of it.
\end{abstract}

\maketitle

\section{\label{sec:intro}Introduction}

Quantum entanglement is at the heart of many quantum information protocols \cite{entanglementreview}, such as quantum state teleportation \cite{teleportation}, and utilised in quantum repeaters \cite{repeaters}, which are fundamental for long-distance quantum communication. Entanglement can also result in Bell nonlocality through the correlations between measurements performed by distant parties, manifested as violations of Bell inequalities \cite{bell,review}. Now this form of nonlocality can be a resource for tasks such as quantum key distribution \cite{ekert,bhk,diqkd}, certifiable randomness expansion \cite{pironio2010random,colbeck,randomnessreview}, delegated quantum computation \cite{ruv}, communication complexity \cite{communication} and measurement-based quantum computation \cite{andersbrowne,hoban}. 

Besides being an information theoretic resource on their own, Bell inequality violations have the remarkable property of witnessing entanglement without the need to know the underlying physical system. In other words, Bell nonlocality witnesses entanglement in the \textit{device-independent} paradigm in which devices are not characterized. But Bell inequality violations can certify more than the mere presence of entanglement and, in fact, they are also useful in the context of \textit{quantum state certification}. In quantum state certification, a device claims to produce systems with particular quantum states, and the goal is to have a task that certifies this claim. The certification task for the source depends very much on the assumptions made in a scenario, such as whether measurement devices can be fully characterised and trusted (device-dependent) or not characterised nor trusted at all (device-independent). When it comes to the certification of a source of entangled particles in a completely device-independent manner, certification is based on correlations violating Bell inequalities and is described as \textit{self-testing} \cite{Mayers2004}. This question has gained a lot of attention in recent years~\cite{McKague2014,Coladangelo2017,Jed1,Ivan, STreview}.
A notable trait of self-testing is the inability to recover the exact form of the state, and measurements: the best one can hope is to certify them up to operations which leave the observed probability distributions invariant. Local isometries and complex conjugation are examples of such operations.

On the other side, in the device-dependent scenario where measurement devices are perfectly characterized, a lot is known, e.g. see~\cite{ashley, markham, TakeuchiMorimae} for recent progress in the efficient certification of quantum states. 
In between these two extreme cases one has different relaxations of the device-independent scenario, being sometimes coined as semi-device-independent. This term was originally introduced in~\cite{BP} for the case in which an upper bound on the dimension of the systems is assumed, but we use it here to describe any scenario between the completely device-dependent and device-independent scenarios. For instance, if one assumes a perfect knowledge about one of the two devices, entanglement can be witnessed through correlations displaying Einstein-Podolsky-Rosen (EPR) steering \cite{erwin,wiseman}, which has led to the study of one-sided device-independent quantum information processing \cite{steeringreview,steeringreview2} and quantum certification based on steering~\cite{IvanMatty, Gheorghiu}. Other works have also considered the problem of state certification by assuming a bound on the dimension of the involved quantum systems~\cite{Armin,Farkas}. 

While all these different scenarios differ in the assumptions invoked for the certification, they are all based on the statistics describing an input-output process consisting of classical inputs, labelling choices of measurements or states, and outputs, associated to measurement results. Our work goes beyond this framework and study certification protocols in which the inputs have a quantum nature. In this scenario, each party could individually generate other characterised quantum systems in a trusted way. These characterised quantum systems can then be used as quantum input into an uncharacterised device. This type of certification naturally appears in the context of semi-quantum nonlocal games~\cite{buscemi} but also in quantum information protocols with no classical analogue such as teleportation \cite{teleportation}. It is also relevant in the context of  device-independent quantum certification, as the characterised quantum systems could themselves have been certified separately in a device-independent manner, see for instance~\cite{bowles}. Our main results consist of different new self-testing protocols using quantum inputs.

\section{Frameworks for quantum state certification}

In this section we identify four basic frameworks for quantum state certification in a bipartite setting, corresponding to four forms of device-independence. Throughout this work, as a simplification, we will assume that in every instance the device produces identical and independently distributed (i.i.d.) copies of the same system. Additionally, in all bipartite scenarios the two parties will be referred to as Alice and Bob. 

\subsubsection*{Device-dependent state certification} The first framework accounts for characterised and trusted measurement devices, which can be applied to systems generated by an untrusted and uncharacterised preparation device. State certification can be achieved by \textit{quantum state tomography} \cite{qse}: informationally complete measurements \cite{Prugovecki1977} can be made on the i.i.d. copies of the quantum system. The probabilities of obtaining different measurement outcomes are used to determine the state. For example, if the source produces one-qubit states, an example of an informationally complete set of measurements are those projective measurements associated to the three Pauli operators $\{\sigma_\mathsf{x},\sigma_\mathsf{y}, \sigma_\mathsf{z}\}$. 
The probability to obtain outcome $a$ when measuring $x$-th measurement on the unknown state $\varrho$ is given by
\begin{equation*}
p(a\vert x) = \textrm{Tr}\left(\M_{a\vert x}\varrho\right),
\end{equation*}
where $\M_{a\vert x}$ denotes the measurement element corresponding to the outcome $a$.
The aim of quantum state tomography is to recover the state $\varrho$ from a given set $\{p(a\vert x), \M_{a\vert x}\}_{a\vert x}$.

An analogous procedure can be described to characterise an unknown quantum measurement using a characterised set of quantum states. The set of quantum states is used as a probe and the probabilities of obtaining different measurement outcomes are used to recover the form of the measurement. The set of states sufficient for this process is called a tomographically complete set of states. For a qubit measurement, a tomographically complete set of states are, for example, the eigenstates of the three Pauli operators.

Performing tomography is however not necessary for quantum state certification in the device-dependent setting. For certification we merely wish to prove that \textit{a particular} state is produced, and thus we only need to establish whether it is that state, or not. A solution to this, for a pure state, is to have a projective measurement with that state as one of its outcomes. For entangled states this might require entangled measurements, but there are other approaches not requiring such complicated measurements~\cite{ashley,TakeuchiMorimae}. 

\subsubsection*{Self-testing} The device-independent scenario is that which completely lacks a characterisation of the devices. In this case, Alice's and Bob's devices are treated as black boxes with classical inputs and classical outputs. The corresponding certification task is named self-testing \cite{Mayers2004}. The aim is to recover the entangled state $\ket{\psi}$ only from the probabilities of obtaining different outputs when certain inputs are chosen. Self-testing can only hope to recover a state able to produce a nonlocal  probability distribution (see \cite{Goh}), which means that it cannot be performed on single systems. 
The starting point in every self-testing procedure is the correctness of the Born rule, which allows to calculate the correlation probabilities when unknown measurements $\M_{a\vert x}$ and $\M_{b \vert y}$ are performed on the shared state $\varrho'$:
\begin{equation*}
p(a,b \vert x,y) = \textrm{Tr}\left[\left(\M_{a\vert x} \otimes \M_{b\vert y}\right)\varrho'\right].
\end{equation*}
Since all one has access to is the probabilities, one cannot differentiate between physical set-ups (involving potentially different states and measurements) that give rise to the same probabilities. For instance, self-testing cannot prove that $\varrho$ is exactly equal to $\ket{\psi}$ but it may allow one to prove that the two states are related by a suitable local isometry $\Phi = \Phi_\rA\otimes\Phi_\rB$:
\begin{equation*}
    \Phi(\varrho') = \ketbra{\psi}{\psi}\otimes\varrho_{junk},
\end{equation*}
where $\varrho_{junk}$ represents the state of the uncorrelated degrees of freedom. 

\subsubsection*{One-sided device-independent certification} 
As mentioned, between these two cases there are methods for certification, known as semi-device-independent, based on assumptions on the devices but that do not require a full characterization. Next we illustrate this approach through two well known examples.

A quantum state can be certified in an asymmetric scenario: one party has characterised measurements while the other treats their devices as black boxes. This certification task is clearly between the device-dependent and the device-independent settings and thus it has been introduced as one-sided device-independent self-testing \cite{IvanMatty,Gheorghiu1}. Here it is possible to carry out tomography using trusted measurement devices, but with only classical inputs and outputs for the black-box devices. The part of the state belonging to the party with uncharacterised devices can be recovered only up to local isometries. Only states which do not admit a local hidden state model, i.e. steerable ones, can be self-tested in this way \cite{steeringreview,steeringreview2}.

\subsubsection*{Bounded dimension self-testing} Certification protocols can be based on an assumption of the dimension of the involved systems. An advantage of this approach is that it can be applied to \textit{prepare-and-measure} scenario \cite{Gallego, Armin, Farkas}. Alice prepares systems which are subsequently measured by Bob; the task is based on communication between two parties thus making it different from the other settings certifying entangled states. The central assumption made in such settings is that the system is associated with a Hilbert space of a fixed dimension, but otherwise devices are not characterised. In \cite{Weixu} prepare-and-measure scenarios are used to certify properties of quantum measurements by assuming the bound on the overlap between the states Alice can prepare, instead of bounding the Hilbert state dimension.

\section{Self-testing with quantum inputs}

In all the previous approaches to self-testing, and independently of the assumptions on the devices, the parties feed the devices with classical information, which can label a state preparation of measurement choice, and observe an output, corresponding to a measurement result. In this work, we consider a different framework in which parties can locally prepare some characterised quantum states, which are later treated as inputs to their untrusted measurement devices. Measurement-device-independent (MDI) protocols are examples of this approach, which is becoming increasingly popular in recent years. Firstly, it has been proven that, in this scenario, all entangled states are capable of exhibiting measurement correlations which cannot be simulated with separable states~\cite{buscemi}, see also~\cite{mdi,IPD,RMVLT}. The same approach has been pursued in \cite{tel,isc2}  to clarify the role of entanglement in quantum teleportation protocols. The main goal of this work is to construct self-testing protocols in this scenario.

\begin{figure}
\centering
\includegraphics[width=0.65\columnwidth]{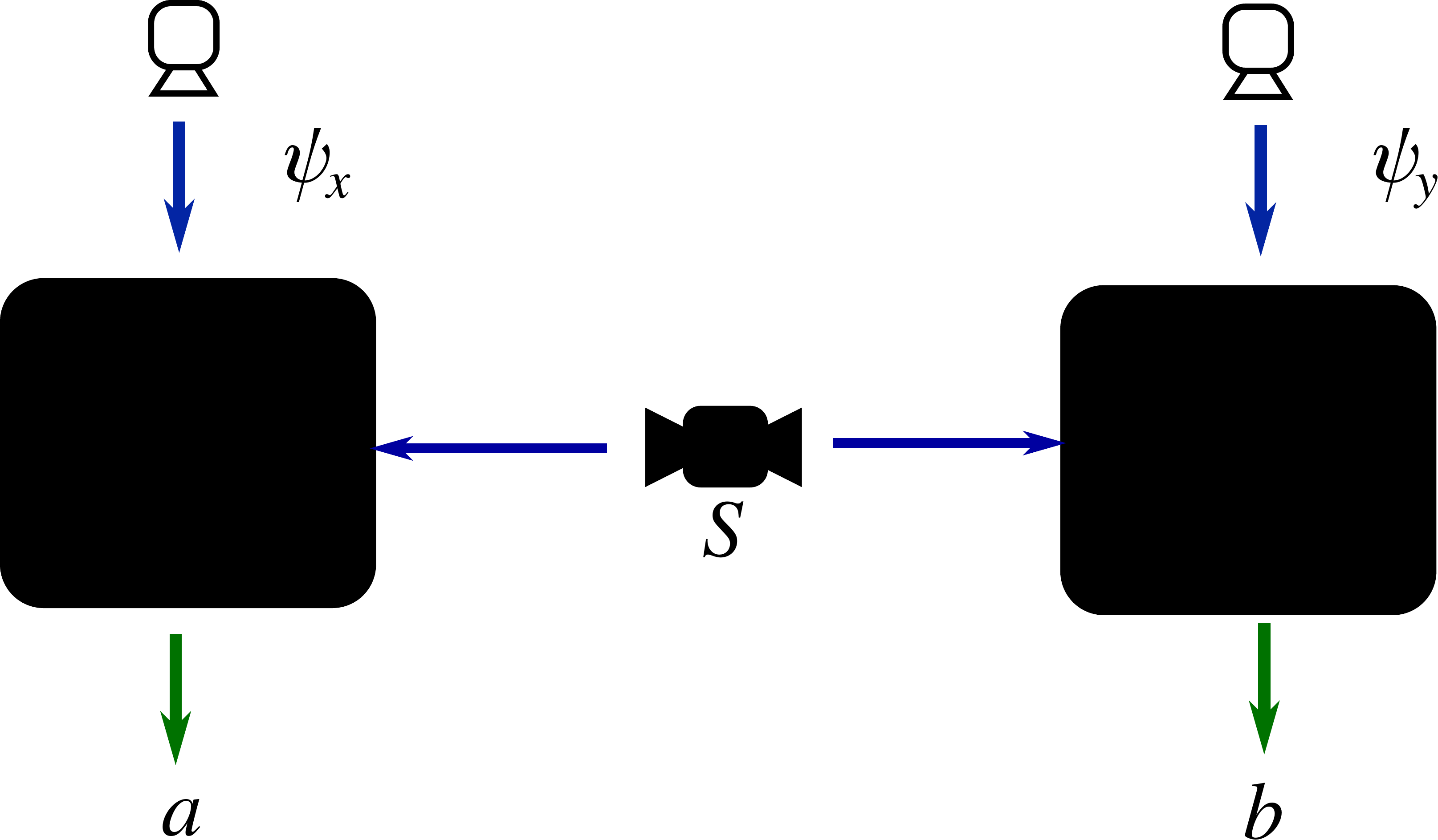}
\caption{Measurement-device-independent scenario: The parties share an unknown state $\varrho$, emitted by the source $S$. The uncharacterised measurement devices receive trusted quantum inputs $\psi_x$ ($\psi_y$). Each party applies a joint measurement on the received quantum input and a share of the state $\varrho$, resulting in the outcomes $a$ and $b$.  \label{fig:qqexample}}
\end{figure}

Before describing self-testing with quantum inputs let us point out what kind of conclusions we can expect. Since, in this scenario, the measurement devices are not trusted along with the source of the systems, they may be associated with Hilbert spaces of arbitrary dimension. Additionally, all of the experimental observations are insensitive to a set of transformations; this is similar to the situation in standard self-testing protocols. So if the underlying experiment deviates from the claimed one in suitable ways as not to alter the observed statistics, these deviations cannot be determined and define an equivalence class of preparations. Any local change of basis to the states and measurements remains hidden, as well as embedding of the state in some Hilbert space of higher or lower dimension. Consequently, the best we can hope for is to find local isometries (one for Alice and one for Bob) relating the state we want to certify with the state shared by Alice and Bob. Importantly, in this scenario, complex conjugation can be dropped from the set of undetectable state transformations. The reason for this is the full characterisation of quantum inputs, which can be chosen from a tomographically complete set of states. Thus we can distinguish the statistics produced by $\ket{\psi}$, in general, from those produced by the state $\ket{\psi^*}$. Similarly to the self-testing nomenclature we call the ideal state \textit{reference state} and the shared state \textit{physical state}. For the sake of simplicity, we restrict our study to protocols where in the ideal scenario parties always apply the Bell state measurement (the projector onto the Bell states of the corresponding dimension). That is, in all experiments with quantum inputs the reference measurement is the Bell state measurement, while the actual measurement the parties apply is named \textit{physical measurement}. Of course, the formalism can in principle be generalized to other measurement settings, but we do not consider them here.

\subsection{Self-testing with only quantum inputs}\label{qqst}

In this section we consider bipartite self-testing in which all parties use quantum inputs, i.e. MDI state certification. The scenario is as follows: two parties, Alice and Bob, share a quantum state ${\varrho'}^{\rA\rB}$. Each of them can perform a joint measurement on their share of the entangled state and the prepared quantum input, $\psi_x^{\rA'}$ for Alice and $\psi_y^{\rB'}$ for Bob. We are using the notation that $\psi\equiv\ket{\psi}\bra{\psi}$ is the projector onto the pure state $\ket{\psi}$. Since the Hilbert spaces are unbounded in dimension, the measurements are modelled as projectors: $\{\M_a^{\rA'\rA}\}_a$ for Alice and $\{\M_b^{\rB\rB'}\}$ for Bob. Measurement outcomes are labelled with $a$ for Alice and $b$ for Bob (see Fig. \ref{fig:qqexample}). The aim of self-testing with quantum inputs is to prove that from the observed statistics $p(a,b\vert \psi_x, \psi_y)$ it follows that there must exist a local isometry transforming  the physical state ${\varrho'}^{\rA\rB}$ into the reference one $\psi^{\rA''\rB''}$. Similarly to standard self-testing we can only hope to certify the presence of pure states. Analogously to the theorem given in \cite{SVZ} we can prove that the correlations of any mixed state can be achieved with a pure state of the same dimension. The proof of this theorem is presented in Appendix \ref{mixed}.

Before stating the main theorem of this section let us recall some of the specificities of the scenario when the parties can prepare tomographically complete set of inputs (for more details see \cite{mdi,IPD}). The observed probabilities can be written in the following way:
\begin{equation*}
p(a,b \vert \psi_x,\psi_y) = \textrm{Tr}\left[{\tilde{\M}_{a,b}}^{\rA'\rB'}\left(\psi_x^{\rA'}\otimes \psi_y^{\rB'}\right)\right],
\end{equation*}
where 
\begin{equation}\label{EffMeas}
{\tilde{\M}_{a,b}}^{\rA'\rB'} = \textrm{Tr}_{\rA\rB}\left[\left(\M_a^{\rA'\rA}\otimes \M_b^{\rB\rB'}\right)\left(\mathds{1}^{\rA'}\otimes {\varrho'}^{\rA\rB}\otimes \mathds{1}^{\rB'}\right)\right]
\end{equation}
is named the \textit{effective measurement}. 
If the set of quantum input states is tomographically complete, in the sense that it is sufficient for quantum process tomography, one can recover the exact form of the effective measurements from the observed probabilities. This insight is in the core of the proof that quantum inputs can successfully probe every entangled state \cite{buscemi,mdi} and its analogue is the central object in the contributions to the understanding of non-classical quantum state teleportation \cite{tel,isc2}. To briefly summarise, if the effective measurement is not a separable operator for every pair $a,b$ the shared state must be entangled. 

The following theorem will identify precisely how the resemblance between the effective measurement and the shared state can be used for the recovery of the state. In particular, if the effective measurements are pure and entangled the self-testing statement for the shared state can be formulated. To state the theorem, we need to introduce some notation. The $d$-dimensional generalized $Z$ and $X$ operators are defined as $Z = \sum_{j=0}^{d-1}\omega^j \ketbra{j}{j}$ and $X = \sum_{j=0}^{d-1}\ketbra{j+1 \mod d}{j}$, respectively, where $\omega = \exp{2\pi i/d}$. These matrices can be used to define an orthonormal basis of qudit Bell states $\ket{\psi_{kl}} = X^kZ^l\ket{\phi^+}$, where $\ket{\phi^+} = \sum_{j=0}^{d-1}\ket{jj}$. As Alice's and Bob's reference measurements are $\{\ketbra{\psi_{kl}}{\psi_{kl}}\}$ their outputs $a$ and $b$ are comprised of two dits $k$ and $l$. 
\begin{thm}
\textit{Let two parties, Alice and Bob, share  the state ${\varrho'}^\emph{\rA\rB}$ and have access to a tomographically complete set of inputs $\{\psi_x\}_x$ and $\{\psi_y\}_y$ respectively. Each party performs a joint measurement on their share of ${\varrho'}^\emph{\rA\rB}$ and quantum input $\psi_{x}$ or $\psi_{y}$. If the correlation probabilities can be written in the form
\begin{eqnarray*}
p\left(a,b \vert \psi_x,\psi_y\right) = \emph{Tr}\left[{\tilde{\M}_{a,b}}^{\emph{\rA}'\emph{\rB}'}\left(\psi_x^{\emph{\rA}'}\otimes \psi_y^{\emph{\rB}'}\right)\right],  \quad \forall a,b,x,y;
\end{eqnarray*}
and $\tilde{\M}^{\emph{\rA}'\emph{\rB}'}_{a,b}$ are such that
\begin{eqnarray}\label{stcondition}
\frac{1}{d^2}\ketbra{\psi}{\psi} &=& (U_a\otimes U_b){(\tilde{\M}_{a,b}^{\emph{\rA}'\emph{\rB}'})}^T(U_a^{\dagger}\otimes U_b^{\dagger})  \quad \forall a,b,
\end{eqnarray}
where $U_a$ and $U_b$ are the correcting unitaries defined as $U_m = \sum_{kl}X^kZ^l\delta_{m,kl}$, 
%
%
then there exists a local isometry $\Phi$ such that 
\begin{equation}\label{statement}
\Phi({{\varrho'}^\emph{\rA\rB}}) = \ketbra{\psi}{\psi}^{\emph{\rA}''\emph{\rB}''}\otimes \varrho_{junk}^{\emph{\rA\rA}'\emph{\rB\rB}'}
\end{equation}}
\qed
\end{thm}

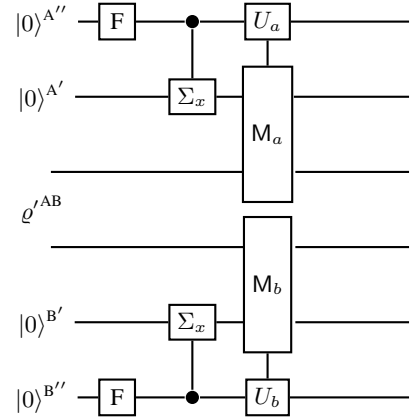
\begin{figure}[h!]
  \centerline{
    \begin{tikzpicture}[thick]
    %
    \tikzstyle{operator} = [draw,fill=white,minimum size=1.5em] 
    \tikzstyle{operator2} = [draw,fill=white,minimum height=1.8cm]
    \tikzstyle{phase} = [fill,shape=circle,minimum size=5pt,inner sep=0pt]
    \tikzstyle{circlewc} = [draw,minimum width=0.3cm]
    %
    \node at (0,0) (q1) {$\ket{0}^{\rA''}$};
    \node at (0,-1) (q2) {$\ket{0}^{\rA'}$};
    \node at (0,-2.5) (q3) {${\varrho'}^{\rA\rB}$};
    \node at (0, -2) (qex1) {};
    \node at (0, -3) (qex2) {};
    \node at (0,-4) (q4) {$\ket{0}^{\rB'}$};
    \node at (0,-5) (q5) {$\ket{0}^{\rB''}$};
    \node at (3.23, -1) (qex3) {};
    \node at (3.25, -2) (qex4) {};
    \node at (2.80, -1) (qex5) {};
    \node at (2.80, -2) (qex6) {};
    \node at (2.80, -3) (qex7) {};
    \node at (3.25, -3) (qex8) {};
    \node at (2.80, -4) (qex9) {};
    \node at (3.25, -4) (qex10) {};
    %
    \node[operator] (op11) at (1,0) {F} edge [-] (q1);
    \node[operator] (op51) at (1,-5) {F} edge [-] (q5);
    %
    \node[phase] (phase11) at (2,0) {} edge [-] (op11);
    \node[circlewc] (circlewc12) at (2,-1) {$\Sigma_x$} edge [-] (q2);
    \draw[-] (phase11) -- (circlewc12);
    \node[circlewc] (circlewc42) at (2,-4) {$\Sigma_x$} edge [-] (q4);
    \node[phase] (phase51) at (2,-5) {} edge [-] (op51);
    \draw[-] (phase51) -- (circlewc42);
    \draw[-] (circlewc12) -- (qex5);
    \draw[-] (qex1)-- (qex6);
    \draw[-] (qex2)-- (qex7);
    \draw[-] (circlewc42)-- (qex9);
    %
    \node[operator] (op31) at (3,0) {$U_a$} edge [-] (phase11);
    \node[operator2] (operator32) at (3, -1.5) {$\M_a$}; 
    \node[operator2] (operator33) at (3, -3.5) {$\M_b$};
    \node[operator] (op35) at (3,-5) {$U_b$} edge [-] (phase51);
    \draw[-] (op31) -- (operator32);
    \draw[-] (op35) -- (operator33);
    %
    \node (end1) at (5,0) {} edge [-] (op31);
    \node (end2) at (5,-1) {} edge [-] (qex3);
    \node (end3) at (5,-2) {} edge [-] (qex4);
    \node (end4) at (5,-3) {} edge [-] (qex8);
    \node (end5) at (5,-4) {} edge [-] (qex10);
    \node (end6) at (5,-5) {} edge [-] (op35);
    \end{tikzpicture}
  }
  \caption{
   Representation of the isometry $\Phi$. It takes as an input the state ${\varrho'}^{\rA\rB}$ and each party performs a unitary operation $U_{a/b}$ conditioned on the outcome of the measurement $\M_{a/b}$. $F$ is the Fourier transform gate acting as $F\ket{j} = \sum_k e^{ijk\pi/d}\ket{k}$, while the second gate is a generalized CNOT gate acting as $CNOT\ket{j}\ket{k} = \ket{j}\ket{j+k}$.
  }
  \label{iso}
\end{figure}

The proof of the theorem is given in Appendix \ref{proof1}. Here we explicitly show the isometry which is used to prove the theorem. The isometry is given in Fig. \ref{iso}. It implicitly assumes that the measurement operators are projective. Since the dimension of the shared state is not assumed, the Naimark extension can be used: if the measurements $\{\M_{a}\}$, $\{\M_{b}\}$ are not projective one can always increase the dimension of the registers $\rA$, $\rB$ and see the measurements as projective on a higher dimensional system. 

The true power of quantum inputs is exhibited when one is interested in the robustness of the self-testing procedure. The standard task of robust self-testing can be phrased as follows: if the conditions for self-testing are approximately met can we still say something about the distance between $\Phi(\varrho)$ and $\psi$? When the set of quantum inputs is tomographically complete the state of the registers $\rA''$ and $\rB''$ of the isometry on Fig. \ref{iso}  can be recovered even if the conditions for the ideal self-testing \eqref{stcondition} are not satisfied. In that case the fidelity between $\tr_{\rA\rA'\rB\rB'}\Phi(\varrho^{\rA\rB})$ and the reference state $\ket{\psi}$ can be directly estimated. Furthermore, if the parties used exactly the Bell state measurement the physical state will be exactly mapped to the state of registers $\rA''$ and $\rB''$, allowing to obtain the tight bound on the fidelity between $\tr_{\rA\rA'\rB\rB'}\Phi(\varrho^{\rA\rB})$ and  $\ket{\psi}$. A noisy Bell state measurement will give only a lower bound on the fidelity. For example, if the noisy Bell state measurements of the form $\M_i' = 0.95B_i + 0.05\mathbb{1}/4$, where $\{B_i\}_i$ is the ideal Bell state measurement, is applied on the state $\varrho = {\phi^+}$, the recovered fidelity between  $\tr_{\rA\rA'\rB\rB'}\Phi(\varrho^{\rA\rB})$ and ${\phi^+}$ will be $0.893$.

Note that the ability to prepare quantum inputs is strictly more than what one can do in a device-independent scenario. Thus, one would expect that whenever there is a standard device-independent self-test for a state $\ket{\psi}$, it can be also performed with quantum inputs. The idea is simple: if some projectors are used to produce measurement correlations which are self-testing the state $\ket{\psi}$, they can be effectively prepared by performing a Bell state measurement and a suitable input. An example of adapting the self-test from the CHSH inequality to the scenario with quantum inputs is provided in Appendix \ref{CHSH}.

The similar overall reasoning about self-testing with only quantum inputs can be applied to every multipartite entangled state. The more detailed discussion is given in Appendix \ref{mltp}. Here we just state the corollary:

\begin{cor}
Self-testing with quantum inputs can recover any pure genuinely multipartite entangled state.
\end{cor}

\subsection{Self-testing with quantum-classical inputs}\label{qcst}

\begin{figure}
\centering
\includegraphics[width=0.65\columnwidth]{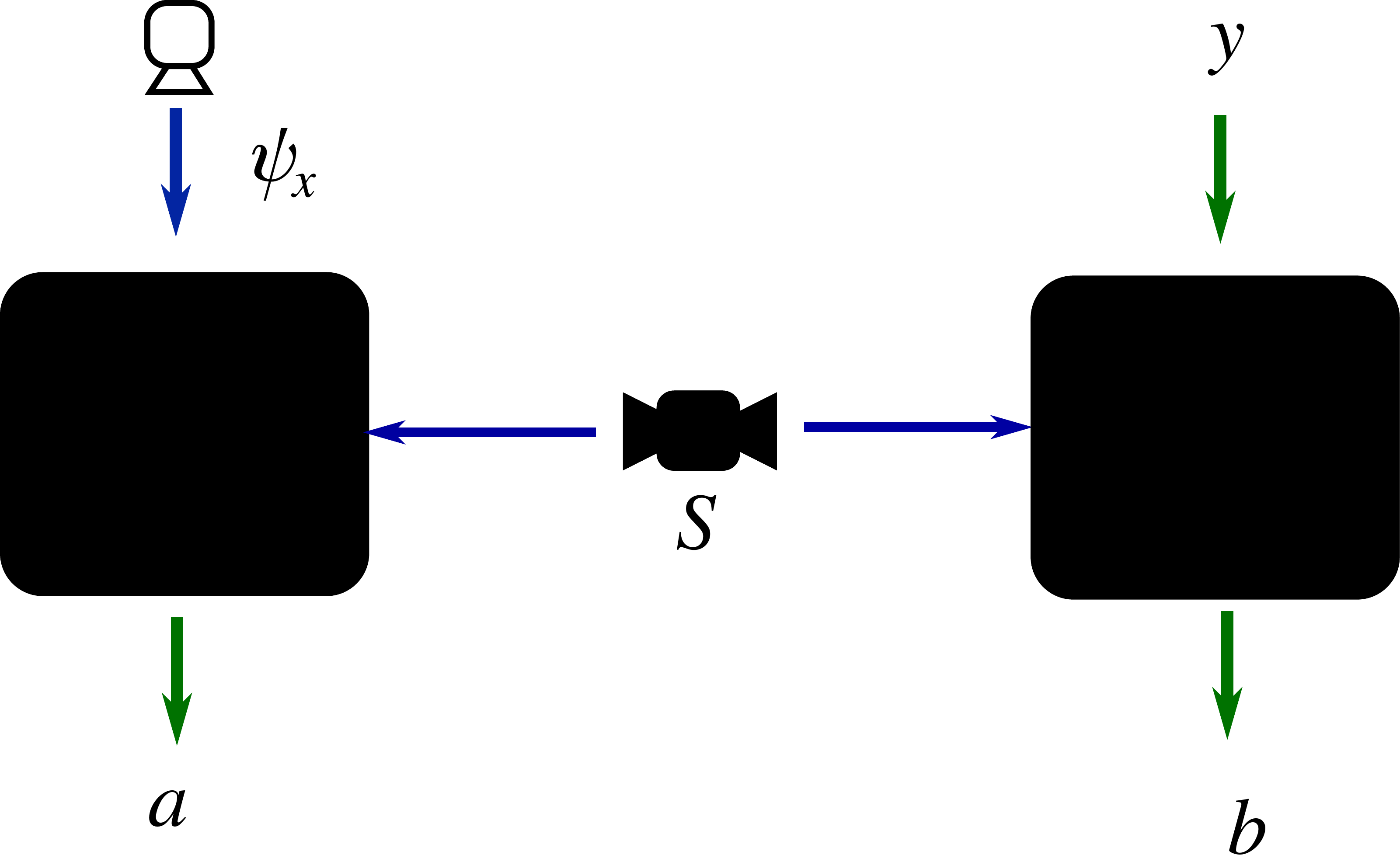}
\caption{Self-testing with quantum-classical inputs: Alice can prepare quantum inputs $\psi_x$ and by measuring them together with a share of a state emitted by the source $S$, she obtains measurement outcome $a$. Bob, on the other side, treats all his devices as black boxes. He labels his measurement choice with a classical input $y$ and obtains the measurement outcome $b$. \label{fig:example}}
\end{figure}

In this section we consider a hybrid scenario in which one party, say Alice, uses quantum inputs, while the other one, Bob, uses classical inputs (see Fig. \ref{fig:example}).  Let us consider the following scenario: Alice and Bob share a state ${\varrho'}^{\rA\rB}$. Alice can prepare quantum inputs $\{\psi_x^{\rA'}, \bar{\psi}_x^{\rA'}\}_x$, where $\bar{\psi_x} = \mathds{1}-\psi_x$, and apply  a joint measurement $\{\M_a^{\rA'\rA}\}_a$, while Bob queries his device with classical input $y$, which corresponds to applying a projective measurement $\{\M_{b|y}\}_b$. In this scenario the probability to obtain outcomes $a$ and $b$, when Alice's quantum input is $\psi_x$ and Bob's classical input is $y$ is
\begin{equation}
    p(a,b|\psi_x,y) = \textrm{Tr}\left[\left(\M_a^{\rA'\rA}\otimes \M_{b|y}\right)\left(\psi_x^{\rA'}\otimes {\varrho'}^{\rA\rB}
    \right)\right].
\end{equation}
For each classical input $y$ we can define the effective measurements
\begin{equation}\label{effmqc}
    \tilde{\M}_{a,b|y}^{\rA'} = \textrm{Tr}_{\rA\rB}\left[\left(\M_a^{\rA'\rA}\otimes \M_{b|y}^\rB\right)\left(\mathds{1}^{\rA'}\otimes{\varrho'}^{\rA\rB}\right)\right].
\end{equation}
Now we are ready to state the theorem which self-tests the state $\ket{\phi^+}$ from a Bell-like expression. Let $\mathcal{I}_{qc}$ be defined in the following way:
\begin{widetext}
\begin{equation}\label{iqc}
\begin{split}
\mathcal{I}_{qc} =  \sum_{a=0,1}(p(a,0|\psi_0,0)+p(a,1|\bar{\psi}_0,0))+\sum_{a=2,3}(p(a,1|\psi_0,0)+p(a,0|\bar{\psi}_0,0) )+ \\
 + 
 \sum_{a=0,2}(p(a,0|\psi_1,1)+p(a,0|\bar{\psi}_1,1))+\sum_{a=1,3}(p(a,1|\psi_1,1)+p(a,0|\bar{\psi}_1,1) ).
 \end{split}
\end{equation}
\end{widetext}
The algebraic maximum of $\mathcal{I}_{qc}$ is equal to $4$. We show that there exist quantum inputs for which the only way to achieve the algebraic maximum is to share a maximally entangled pair of qubits. 
\begin{thm}\label{statementqc}
\textit{Let two parties, Alice and Bob, share a  state ${\varrho'}^{\rA\rB}$. Furthermore, let Alice use quantum inputs $\psi_0 = \ketbra{0}{0}$, $\bar{\psi}_0 = \ketbra{1}{1}$, $\psi_1 = \ketbra{+}{+}$ and $\bar{\psi}_1 = \ketbra{-}{-}$. If they observe $\mathcal{I}_{qc} = 4$ where $\mathcal{I}_{qc}$ is defined in \eqref{iqc}
then there exists a local isometry $\Phi$ such that 
\begin{equation}\label{isomeqc}
\Phi_{qc}({\varrho'}^{\rA\rB}) = \ketbra{\phi^+}{\phi^+}\otimes \varrho_{junk}
\end{equation}}
\qed
\end{thm}
The detailed proof is given in Appendix \ref{proof3}. Here we give the intuition for the proof. The main insight comes from the observation that the algebraic maximum of $\mathcal{I}_{qc}$ implies that the effective measurements \eqref{effmqc} can be exactly recovered. Once they are recovered, one can use methods from standard and one-sided-device independent self-testing to prove that a convenient isometry transforms ${\varrho'}^{\rA\rB}$ into $\phi^+$. The isometry is explicitly given in Fig. \ref{isoQC}. 
Operators $\mathbb{M_z}$ and $\mathbb{M_x}$ are given as
\begin{align*}
    \mathbb{M_z} &= \M_0 + \M_1 - \M_2 - \M_3;\\
    \mathbb{M_x} &= \M_0 - \M_1 + \M_2 -\M_3.
\end{align*}

\begin{figure}
  \centerline{
    \begin{tikzpicture}[thick]
    %
    \tikzstyle{operator} = [draw,fill=white,minimum size=1.5em] 
    \tikzstyle{operator2} = [draw,fill=white,minimum height=1.8cm]
    \tikzstyle{phase} = [fill,shape=circle,minimum size=5pt,inner sep=0pt]
    \tikzstyle{circlewc} = [draw,minimum width=0.3cm]
    %
    \node at (0,0) (q1) {$\ket{0}^{\rA''}$};
    \node at (0,-1) (q2) {$\ket{0}^{\rA_1'}$};
    \node at (0,-2.5) (q3) {${\varrho'}^{\rA\rB}$};
    \node at (0, -2) (qex1) {};
    \node at (0, -3) (qex2) {};
    \node at (0,-4) (q4) {$\ket{0}^{\rB'}$};
    \node at (1.8, -1) (qex3) {};
    \node at (1.8, -2) (qex4) {};
    \node at (2.2,-1) (qq1) {};
    \node at (2.9,-1) (qq2) {};
    \node at (4,-1) (qq3) {$\ket{+}^{\rA_2'}$};
    \node at (4.8,-1) (qq4) {};
    \node at (2.2,-2) (qq5) {};
    \node at (4.8,-2) (qq6) {};
    \node at (5.2, -1) (qq7) {};
    \node at (5.2, -2) (qq8) {};
    %
    \node[operator] (op11) at (1,0) {$H$} edge [-] (q1);
    \node[operator] (op41) at (1,-4) {$H$} edge [-] (q4);
    %
    \node[phase] (phase11) at (2,0) {} edge [-] (op11);
    \node[operator2] (circlewc12) at (2,-1.5) {$\mathbb{M_z}$};
    \draw[-] (phase11) -- (circlewc12);
    \draw[-] (qex3) -- (q2);
    \draw[-] (qex4) -- (qex1);
    \node[phase] (phase42) at (2,-4) {} edge [-] (op41);
    \node[circlewc] (circlewc13) at (2,-3) {$\sigma_{\mathsf{z}}$} edge [-] (qex2);
    \draw[-] (phase42) -- (circlewc13);
    \draw[-] (qq2) -- (qq1);
    \node[operator] (op12) at (3.5,0) {$H$} edge [-] (phase11);
    \node[operator] (op42) at (3.5,-4) {$H$} edge [-] (phase42);
    \node[phase] (phase12) at (5,0) {} edge [-] (op12);
    \node[operator2] (circlewc22) at (5,-1.5) {$\mathbb{M_x}$};
    \draw[-] (phase12) -- (circlewc22);
    \draw[-] (qq4) -- (qq3);
    \draw[-] (qq5) -- (qq6);
    \node[phase] (phase42) at (5,-4) {} edge [-] (op42);
    \node[circlewc] (circlewc23) at (5,-3) {$\sigma_{\mathsf{x}}$} edge [-] (circlewc13);
    \draw[-] (phase42) -- (circlewc23);
    \node (end1) at (6,0) {} edge [-] (phase12);
    \node (end2) at (6,-1) {} edge [-] (qq7);
    \node (end3) at (6,-2) {} edge [-] (qq8);
    \node (end4) at (6,-3) {} edge [-] (circlewc23);
    \node (end5) at (6,-4) {} edge [-] (phase42);
    \end{tikzpicture}
  }
  \caption{
   Representation of the local isometry $\Phi_{qc}$. It takes as an input the state ${\varrho'}^{\rA\rB}$ and resembled the standard SWAP isometry. The systems $\rA_1'$ and $\rA_2'$ can be discarded at the end of the process.
  }
  \label{isoQC}
\end{figure}

\section{Basic quantum information protocols as self-tests}

So far we have introduced self-testing in two different semi-device-independent scenarios. In this section we show that, besides being natural extensions of standard self-testing, the introduced protocols have a practical importance in relating self-testing to some of the most widely used quantum information protocols. In section \ref{tlpst} we discuss how quantum state teleportation can be viewed as a self-test, while in section \ref{networkst} we show how one can certify the set of states composing a quantum repeater or a quantum network. 

\subsection{Quantum state teleportation as a self-test}\label{tlpst}

\begin{figure}[h!]
\centering
\includegraphics[width=0.65\columnwidth]{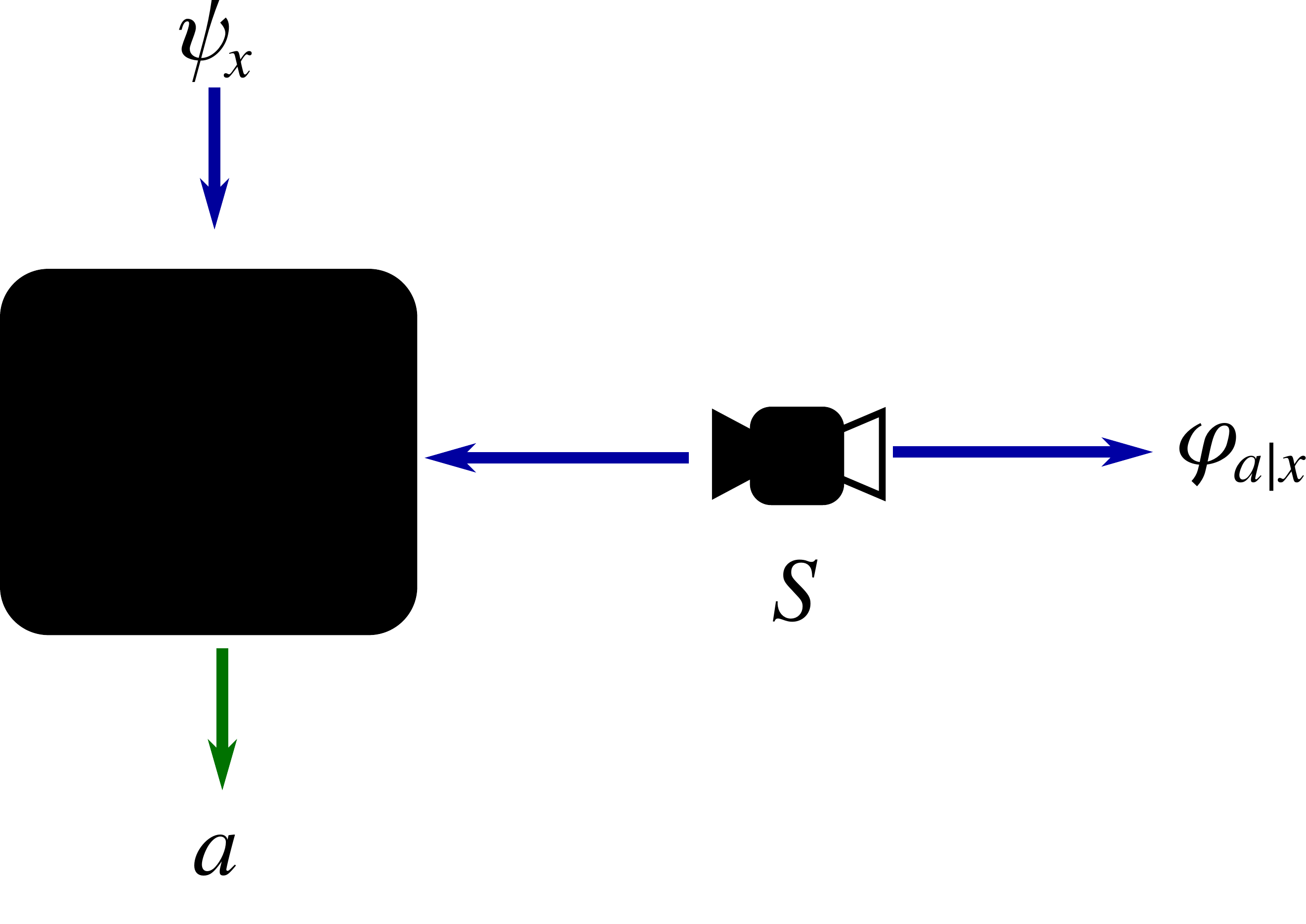}
\caption{Quantum state teleportation: Alice applies a global measurement on the state $\psi_x$ and her share of the state emitted by the source $S$. Bob can apply quantum state tomography and learn exactly his reduced state $\varphi_{a|x}$.\label{fig:tlp}}
\end{figure}

As noted in \cite{tel}, quantum state teleportation can be seen as a representative of one-sided-measurement-device-independent protocols. Indeed, Alice uses a quantum input, performs a joint measurement, while Bob performs quantum state tomography and learns his reduced state. Note that in the spirit of \cite{tel} we do not involve the part of the protocol in which Alice communicates the outcome of her measurement to Bob and he applies the correcting unitary. The correcting unitary can alternatively be applied by a verifier which supervises the teleportation experiment.  The reduced state of Bob $\varphi_{a|x}$ is obtained through the following expression
\begin{equation}
    \varphi_{a|x}= \textrm{Tr}_{\rA'\rA}\left[\left(\M_a^{\rA'\rA}\otimes \mathds{1}^\rB\right)\left(\psi_x^{\rA'}\otimes {\varrho'}^{\rA\rB}\right)\right],
\end{equation}
where $\psi_x^{\rA'}$ is a quantum input, i.e. a state to be teleported, $\M_a^{\rA'\rA}$ is the measurement Alice applies, while ${\varrho'}^{\rA\rB}$ is the state shared between Alice and Bob (see Fig. \ref{fig:tlp}).\\
The success of a teleportation experiment is usually assessed from an average teleportation fidelity, defined as
\begin{equation*}
    \bar{F}_{tel} = \frac{1}{|x|}\sum_{a,x}\frac{\bra{\psi_x}U_a\varphi_{a|x}U_a^{\dagger}\ket{\psi_x}}{p(a|\psi_x)},
\end{equation*}
where $|x|$ is the total number of input states, and $p(a|\psi_x)$ is the probability to obtain outcome $a$ when the input state is $\psi_x$. It was proven in \cite{HHH} that in case the input states are tomographically complete,  the state having Bell state fidelity $F_s(\varrho^{\rA\rB}) = \bra{\phi^+}\varrho^{\rA\rB}\ket{\phi^+}$ leads to the average teleportation fidelity of $(F_s(\varrho^{\rA\rB})d+1)/(d+1)$, where $d$ is the dimension of the states to be teleported. This can be seen also as a self-testing statement: the observed average teleportation fidelity $\bar{F}_{tel}$  gives a lower bound to the Bell state fidelity $F_s$. However, it is obtained under assumption that the shared state is of dimension $d^2$. Here we show how to estimate a lower bound to the Bell state fidelity of the state shared between Alice and Bob from an arbitrary teleportation experiment, including the case when the set of input states is not tomographically complete. \\

As explained in \cite{tel} a teleportation experiment can be characterized by the effective teleportation measurement
\begin{equation*}
\tilde{\M}_{a} = \textrm{Tr}_{\rA\rB}\left[\left(\M_a^{\rA'\rA}\otimes \mathds{1}^\rB\right)\left(\mathds{1}^{\rA'}\otimes {\varrho'}^{\rA\rB}\right)\right].
\end{equation*}
This is clearly tightly related to the effective measurement of Eq.~\eqref{EffMeas}, but now in this new scenario. If the set of input states is tomographically complete, $\tilde{\M}_a$ can be recovered exactly from the set of teleported states $\varphi_{a|x}$. Otherwise, a teleportation experiment is characterized by the set of effective teleportation measurements compatible with the relation
\begin{equation*}
    \varphi_{a|x}^\rB = \textrm{Tr}_{\rA'}\left[\tilde{\M}_a\left(\psi_x^{\rA'}\otimes\mathds{1}^\rB\right)\right].
\end{equation*}
Any set of bipartite operators $\tilde{\N}_a^{\rA'\rB}$ that have a positive partial transposition and  satisfy the no-signalling condition
\begin{equation*}
    \sum_a\tilde{\N}_a^{\rA'\rA} = \mathds{1}\otimes\varrho_r, \qquad \forall a 
\end{equation*}
are valid effective teleportation measurements~\cite{isc2,mattybelen}.

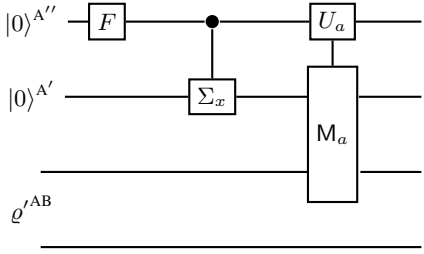
\begin{figure}[h!]
  \centerline{
    \begin{tikzpicture}[thick]
    %
    \tikzstyle{operator} = [draw,fill=white,minimum size=1.5em] 
    \tikzstyle{operator2} = [draw,fill=white,minimum height=1.8cm]
    \tikzstyle{phase} = [fill,shape=circle,minimum size=5pt,inner sep=0pt]
    \tikzstyle{circlewc} = [draw,minimum width=0.3cm]
    %
    \node at (0,0) (q1) {$\ket{0}^{\rA''}$};
    \node at (0,-1) (q2) {$\ket{0}^{\rA'}$};
    \node at (0,-2.5) (q3) {${\varrho'}^{\rA\rB}$};
    \node at (0, -2) (qex1) {};
    \node at (0, -3) (qex2) {};
    \node at (4.23, -1) (qex3) {};
    \node at (4.25, -2) (qex4) {};
    \node at (3.80, -1) (qex5) {};
    \node at (3.80, -2) (qex6) {};
    \draw[-] (qex1)--(qex6);
    \node[operator] (op11) at (1,0) {$F$} edge [-] (q1);
    %
    \node[phase] (phase11) at (2.4,0) {} edge [-] (op11);
    \node[circlewc] (circlewc12) at (2.4,-1) {$\Sigma_x$} edge [-] (q2);
    \draw[-] (phase11) -- (circlewc12);
    %
    \node[operator] (op31) at (4,0) {$U_a$} edge [-] (phase11);
    \node[operator2] (operator32) at (4, -1.5) {$\M_a$}; 
    \draw[-] (op31)--(operator32);
    \draw[-] (circlewc12) -- (qex5);
    \node (end1) at (5.3,0) {} edge [-] (op31);
    \node (end2) at (5.3,-1) {} edge [-] (qex3);
    \node (end3) at (5.3,-2) {} edge [-] (qex4);
    \node (end4) at (5.3,-3) {} edge [-] (qex2);
    %
    \end{tikzpicture}
  }
  \caption{
  Circuit used for self-testing from quantum state teleportation. Alice performs a unitary operation $U_{a}$ conditioned on the outcome of the measurement $\M_{a}$. $F$ is the Fourier transform gate acting as $F\ket{j} = \sum_k e^{ijk\pi/d}\ket{k}$, while the second gate is a generalized CNOT gate acting as $CNOT\ket{j}\ket{k} = \ket{j}\ket{j+k}$.
  }
  \label{isoTel}
\end{figure}

Observe now the quantum circuit  on Fig. \ref{isoTel}. Let us denote the output of the circuit as $\psi_o^{\rA''\rA'\rA\rB}$. In case ${\varrho'}^{\rA\rB}$ is maximally entangled and $\{\M_a\}$ is the  Bell state measurement, the state $\varrho_o = \textrm{Tr}_{\rA\rA'}\psi_o$ is pure and maximally entangled. In fact, since the given quantum circuit is a valid isometry the fidelity between $\varrho_o$ and $\ket{\phi^+}$ lower bounds the fidelity between ${\varrho'}^{\rA\rB}$ and $\ket{\phi^+}$. Since there is no proof that the circuit we use is the optimal isometry, the optimal fidelity might only be higher. In principle, when the set $\{\psi_x\}_x$ is not tomographically complete we cannot know exactly $\varrho_o$. However, since
\begin{equation*}
    \varrho_o = \frac{1}{d}\sum_aU_a^{\rA''}{\tilde{\M}_a}^{T_{\rA''}}{U_a^{\dagger}}^{\rA''}
\end{equation*}
we can optimize over all effective teleportation measurements compatible with the observed teleportation data.

\begin{figure}
    \begin{tikzpicture}[scale=0.95]
      \begin{axis}[
       xlabel = Parameter $p$,
    ylabel = {Fidelity lower bound},
        xmin=0,
        xmax=1,
        xtick={0.25,0.5,0.75,1},
        ymin=0,
        ymax=1,
        ytick={0.25,0.5,0.75,1},
	axis background/.style={fill=white} , set layers, cell picture=true,
	legend pos = north west
        ]
        \addplot [color=orange, draw=orange, thick, mark=none]
        table[row sep=crcr]{
0.005000000000000   0.000000000013509\\
   0.010000000000000   0.000000000019679\\
   0.015000000000000   0.000000000027680\\
   0.020000000000000   0.000000000037689\\
   0.025000000000000   0.000000000049741\\
   0.030000000000000   0.000000000063627\\
   0.035000000000000   0.000000000004518\\
   0.040000000000000   0.000000000005565\\
   0.045000000000000   0.000000000006575\\
   0.050000000000000   0.000000000008055\\
   0.055000000000000   0.000000000009802\\
   0.060000000000000   0.000000000011861\\
   0.065000000000000   0.000000000014284\\
   0.070000000000000   0.000000000017137\\
   0.075000000000000   0.000000000020534\\
   0.080000000000000   0.000000000024553\\
   0.085000000000000   0.000000000029323\\
   0.090000000000000   0.000000000033349\\
   0.095000000000000   0.000000000037870\\
   0.100000000000000   0.000000000042892\\
   0.105000000000000   0.000000000048459\\
   0.110000000000000   0.000000000054614\\
   0.115000000000000   0.000000000061401\\
   0.120000000000000   0.000000000004822\\
   0.125000000000000   0.000000000005467\\
   0.130000000000000   0.000000000006190\\
   0.135000000000000   0.000000000007000\\
   0.140000000000000   0.000000000007910\\
   0.145000000000000   0.000000000008931\\
   0.150000000000000   0.000000000010073\\
   0.155000000000000   0.000000000011351\\
   0.160000000000000   0.000000000012778\\
   0.165000000000000   0.000000000014380\\
   0.170000000000000   0.000000000016178\\
   0.175000000000000   0.000000000018198\\
   0.180000000000000   0.000000000020567\\
   0.185000000000000   0.000000000023364\\
   0.190000000000000   0.000000000026644\\
   0.195000000000000   0.000000000030054\\
   0.200000000000000   0.000000000036735\\
   0.205000000000000   0.000000000044908\\
   0.210000000000000   0.000000000018716\\
   0.215000000000000   0.000000000023197\\
   0.220000000000000   0.000000000028716\\
   0.225000000000000   0.000000000013690\\
   0.230000000000000   0.000000000029854\\
   0.235000000000000   0.000000000010355\\
   0.240000000000000   0.000000000013294\\
   0.245000000000000   0.000000000016839\\
   0.250000000000000   0.000000000007679\\
   0.255000000000000   0.000000000009314\\
   0.260000000000000   0.000000000011072\\
   0.265000000000000   0.000000000004324\\
   0.270000000000000   0.000000000004701\\
   0.275000000000000   0.000000000003330\\
   0.280000000000000   0.000000000005263\\
   0.285000000000000   0.000000000007455\\
   0.290000000000000   0.000000000008342\\
   0.295000000000000   0.000000000010735\\
   0.300000000000000   0.000000000011359\\
   0.305000000000000   0.000000000008821\\
   0.310000000000000   0.000000000011042\\
   0.315000000000000   0.000000000004460\\
   0.320000000000000   0.000000000003859\\
   0.325000000000000   0.000000000013079\\
   0.330000000000000   0.000000000010790\\
   0.335000000000000   0.000000000026324\\
   0.340000000000000   0.000000000020383\\
   0.345000000000000   0.000000000021598\\
   0.350000000000000   0.000000000025332\\
   0.355000000000000   0.000000000008812\\
   0.360000000000000   0.000000000025249\\
   0.365000000000000   0.000000000021603\\
   0.370000000000000   0.000000000008306\\
   0.375000000000000   0.000000000012859\\
   0.380000000000000   0.000000000018145\\
   0.385000000000000   0.000000000006004\\
   0.390000000000000   0.000000000018884\\
   0.395000000000000   0.000000000009602\\
   0.400000000000000   0.000000000005827\\
   0.405000000000000   0.000000000012565\\
   0.410000000000000   0.000000000003880\\
   0.415000000000000   0.000000000003406\\
   0.420000000000000   0.000136854061819\\
   0.425000000000000   0.001274351159820\\
   0.430000000000000   0.003396487327665\\
   0.435000000000000   0.006285190806966\\
   0.440000000000000   0.009774872296010\\
   0.445000000000000   0.013749316200720\\
   0.450000000000000   0.018125808230921\\
   0.455000000000000   0.022843626269081\\
   0.460000000000000   0.027856795960282\\
   0.465000000000000   0.033129556788034\\
   0.470000000000000   0.038633444615957\\
   0.475000000000000   0.044345348932794\\
   0.480000000000000   0.050246194261020\\
   0.485000000000000   0.056320001450171\\
   0.490000000000000   0.062553228169101\\
   0.495000000000000   0.068934254416895\\
   0.500000000000000   0.075453022941037\\
   0.505000000000000   0.082100755298088\\
   0.510000000000000   0.088869715466612\\
   0.515000000000000   0.095753055429948\\
   0.520000000000000   0.102744675163333\\
   0.525000000000000   0.109839094437420\\
   0.530000000000000   0.117031369852776\\
   0.535000000000000   0.124317033464728\\
   0.540000000000000   0.131692010898015\\
   0.545000000000000   0.139152583131122\\
   0.550000000000000   0.146695338003496\\
   0.555000000000000   0.154317134466517\\
   0.560000000000000   0.162015071323870\\
   0.565000000000000   0.169786460510353\\
   0.570000000000000   0.177628798349905\\
   0.575000000000000   0.185539769212057\\
   0.580000000000000   0.193517191945430\\
   0.585000000000000   0.201559029541297\\
   0.590000000000000   0.209663369414549\\
   0.595000000000000   0.217828411818738\\
   0.600000000000000   0.226052459459896\\
   0.605000000000000   0.234333908521738\\
   0.610000000000000   0.242671240576850\\
   0.615000000000000   0.251062991295893\\
   0.620000000000000   0.259507774787890\\
   0.625000000000000   0.268004298705698\\
   0.630000000000000   0.276551315680039\\
   0.635000000000000   0.285147647673090\\
   0.640000000000000   0.293792164379353\\
   0.645000000000000   0.302483778118038\\
   0.650000000000000   0.311221463448553\\
   0.655000000000000   0.320004219466136\\
   0.660000000000000   0.328831089596113\\
   0.665000000000000   0.337701153451237\\
   0.670000000000000   0.346613524939559\\
   0.675000000000000   0.355567349494486\\
   0.680000000000000   0.364561802511830\\
   0.685000000000000   0.373596087568108\\
   0.690000000000000   0.382669434740030\\
   0.695000000000000   0.391781099206609\\
   0.700000000000000   0.400930358024642\\
   0.705000000000000   0.410116509158850\\
   0.710000000000000   0.419338885339234\\
   0.715000000000000   0.428596824066515\\
   0.720000000000000   0.437889675223402\\
   0.725000000000000   0.447216823112724\\
   0.730000000000000   0.456577663867342\\
   0.735000000000000   0.465971632470077\\
   0.740000000000000   0.475399232125578\\
   0.745000000000000   0.484860429252252\\
   0.750000000000000   0.494354574620444\\
   0.755000000000000   0.503881022294369\\
   0.760000000000000   0.513439138467651\\
   0.765000000000000   0.523028302803142\\
   0.770000000000000   0.532647908825355\\
   0.775000000000000   0.542297363684921\\
   0.780000000000000   0.551976088499954\\
   0.785000000000000   0.561683516918596\\
   0.790000000000000   0.571419091606193\\
   0.795000000000000   0.581182280387786\\
   0.800000000000000   0.590972550762368\\
   0.805000000000000   0.600789389848441\\
   0.810000000000000   0.610632292415836\\
   0.815000000000000   0.620500773349331\\
   0.820000000000000   0.630394338469655\\
   0.825000000000000   0.640312524836428\\
   0.830000000000000   0.650254874061795\\
   0.835000000000000   0.660220933700180\\
   0.840000000000000   0.670210281876318\\
   0.845000000000000   0.680222473623537\\
   0.850000000000000   0.690257097578539\\
   0.855000000000000   0.700313739436115\\
   0.860000000000000   0.710392012732237\\
   0.865000000000000   0.720491522064751\\
   0.870000000000000   0.730611887354268\\
   0.875000000000000   0.740752736856064\\
   0.880000000000000   0.750913707209560\\
   0.885000000000000   0.761094443476780\\
   0.890000000000000   0.771294598542450\\
   0.895000000000000   0.781513834210114\\
   0.900000000000000   0.791751818836561\\
   0.905000000000000   0.802008230535257\\
   0.910000000000000   0.812282749552464\\
   0.915000000000000   0.822575073172748\\
   0.920000000000000   0.832884890606011\\
   0.925000000000000   0.843211911151237\\
   0.930000000000000   0.853555846291090\\
   0.935000000000000   0.863916413801063\\
   0.940000000000000   0.874293338334953\\
   0.945000000000000   0.884686350242337\\
   0.950000000000000   0.895095185743335\\
   0.955000000000000   0.905519587322566\\
   0.960000000000000   0.915959314587421\\
   0.965000000000000   0.926414100286032\\
   0.970000000000000   0.936883716805291\\
   0.975000000000000   0.947367934725774\\
   0.980000000000000   0.957866512485524\\
   0.985000000000000   0.968379238870826\\
   0.990000000000000   0.978905872897904\\
   0.995000000000000   0.989446195015420\\
   1.000000000000000   0.999999933065841\\};
\addlegendentry{Case 1}
        \addplot[color=magenta, draw=magenta, very thick, dashed, mark=none]
        table[row sep=crcr]{
0.005000000000000   0.004444392587237\\
   0.010000000000000   0.008888837287583\\
   0.015000000000000   0.013333274235392\\
   0.020000000000000   0.017777750262956\\
   0.025000000000000   0.022222197880019\\
   0.030000000000000   0.026666591052165\\
   0.035000000000000   0.031111088991483\\
   0.040000000000000   0.035555520976593\\
   0.045000000000000   0.039999941668184\\
   0.050000000000000   0.044444419169065\\
   0.055000000000000   0.048888852121841\\
   0.060000000000000   0.053333293062821\\
   0.065000000000000   0.057777751832268\\
   0.070000000000000   0.062222183759823\\
   0.075000000000000   0.066666637812513\\
   0.080000000000000   0.071111088630082\\
   0.085000000000000   0.075555479341681\\
   0.090000000000000   0.079999980756879\\
   0.095000000000000   0.084444424523781\\
   0.100000000000000   0.088888868715001\\
   0.105000000000000   0.093333313280639\\
   0.110000000000000   0.097777758018932\\
   0.115000000000000   0.102222202618535\\
   0.120000000000000   0.106666647691972\\
   0.125000000000000   0.111111092825857\\
   0.130000000000000   0.115555538251860\\
   0.135000000000000   0.119999983823869\\
   0.140000000000000   0.124444369251322\\
   0.145000000000000   0.128888820123389\\
   0.150000000000000   0.133333271482478\\
   0.155000000000000   0.137777721165459\\
   0.160000000000000   0.142222169612613\\
   0.165000000000000   0.146666616257216\\
   0.170000000000000   0.151111061298658\\
   0.175000000000000   0.155555504558769\\
   0.180000000000000   0.159999982599913\\
   0.185000000000000   0.164444423854121\\
   0.190000000000000   0.168888864345493\\
   0.195000000000000   0.173333303852657\\
   0.200000000000000   0.177777742103437\\
   0.205000000000000   0.182222178944538\\
   0.210000000000000   0.186666614039484\\
   0.215000000000000   0.191111047268786\\
   0.220000000000000   0.195555537577014\\
   0.225000000000000   0.199999977988231\\
   0.230000000000000   0.204444418351163\\
   0.235000000000000   0.208888862209752\\
   0.240000000000000   0.213333303813796\\
   0.245000000000000   0.217777745198766\\
   0.250000000000000   0.222222186539648\\
   0.255000000000000   0.226666627910468\\
   0.260000000000000   0.231111069245996\\
   0.265000000000000   0.235555508609757\\
   0.270000000000000   0.239999950568892\\
   0.275000000000000   0.244444392957049\\
   0.280000000000000   0.248888835813704\\
   0.285000000000000   0.253333279172676\\
   0.290000000000000   0.257777723098868\\
   0.295000000000000   0.262222167623568\\
   0.300000000000000   0.266666612886939\\
   0.305000000000000   0.271111058911232\\
   0.310000000000000   0.275555505713063\\
   0.315000000000000   0.279999953240254\\
   0.320000000000000   0.284444401538433\\
   0.325000000000000   0.288888853698963\\
   0.330000000000000   0.293333308516146\\
   0.335000000000000   0.297777761181615\\
   0.340000000000000   0.302222160341535\\
   0.345000000000000   0.306666626659683\\
   0.350000000000000   0.311111098402604\\
   0.355000000000000   0.315555547261787\\
   0.360000000000000   0.319999959975218\\
   0.365000000000000   0.324444417269431\\
   0.370000000000000   0.328888876551784\\
   0.375000000000000   0.333333323066932\\
   0.380000000000000   0.337777768620960\\
   0.385000000000000   0.342222213745429\\
   0.390000000000000   0.346666658767801\\
   0.395000000000000   0.351111103550319\\
   0.400000000000000   0.355555473900080\\
   0.405000000000000   0.359999921124890\\
   0.410000000000000   0.364444368378319\\
   0.415000000000000   0.368888814163006\\
   0.420000000000000   0.373333259643604\\
   0.425000000000000   0.377777704874715\\
   0.430000000000000   0.382222149561062\\
   0.435000000000000   0.386666593689322\\
   0.440000000000000   0.391111038230901\\
   0.445000000000000   0.395555467999316\\
   0.450000000000000   0.399999990902902\\
   0.455000000000000   0.404444358963721\\
   0.460000000000000   0.408888869605342\\
   0.465000000000000   0.413333292888104\\
   0.470000000000000   0.417777717054605\\
   0.475000000000000   0.422222163121389\\
   0.480000000000000   0.426666603494970\\
   0.485000000000000   0.431111067758841\\
   0.490000000000000   0.435555512517188\\
   0.495000000000000   0.439999931616184\\
   0.500000000000000   0.444444333192117\\
   0.505000000000000   0.448899973458609\\
   0.510000000000000   0.453377723917542\\
   0.515000000000000   0.457877713754532\\
   0.520000000000000   0.462399969051399\\
   0.525000000000000   0.466944393739931\\
   0.530000000000000   0.471511046755197\\
   0.535000000000000   0.476099932121882\\
   0.540000000000000   0.480711046377896\\
   0.545000000000000   0.485344385975368\\
   0.550000000000000   0.489999949109969\\
   0.555000000000000   0.494677735250922\\
   0.560000000000000   0.499377743757099\\
   0.565000000000000   0.504099973612989\\
   0.570000000000000   0.508844353760882\\
   0.575000000000000   0.513611034089132\\
   0.580000000000000   0.518399933175960\\
   0.585000000000000   0.523211050802983\\
   0.590000000000000   0.528044388302490\\
   0.595000000000000   0.532899946366044\\
   0.600000000000000   0.537777725219501\\
   0.605000000000000   0.542677725769431\\
   0.610000000000000   0.547599950488620\\
   0.615000000000000   0.552544397961728\\
   0.620000000000000   0.557511066858340\\
   0.625000000000000   0.562499959477338\\
   0.630000000000000   0.567511068189811\\
   0.635000000000000   0.572544394845347\\
   0.640000000000000   0.577599937612461\\
   0.645000000000000   0.582677743084795\\
   0.650000000000000   0.587777739145228\\
   0.655000000000000   0.592899946510819\\
   0.660000000000000   0.598044386910866\\
   0.665000000000000   0.603211050202568\\
   0.670000000000000   0.608399936437480\\
   0.675000000000000   0.613611046062196\\
   0.680000000000000   0.618844379376498\\
   0.685000000000000   0.624099935972993\\
   0.690000000000000   0.629377713749030\\
   0.695000000000000   0.634677727004551\\
   0.700000000000000   0.639999945905818\\
   0.705000000000000   0.645344390035964\\
   0.710000000000000   0.650711060425538\\
   0.715000000000000   0.656099954202282\\
   0.720000000000000   0.661511071839180\\
   0.725000000000000   0.666944354889390\\
   0.730000000000000   0.672399964599130\\
   0.735000000000000   0.677877689367647\\
   0.740000000000000   0.683377728017735\\
   0.745000000000000   0.688899911922024\\
   0.750000000000000   0.694444426786138\\
   0.755000000000000   0.700011088158525\\
   0.760000000000000   0.705599912789182\\
   0.765000000000000   0.711211037701564\\
   0.770000000000000   0.716844487323889\\
   0.775000000000000   0.722501245728477\\
   0.780000000000000   0.728181443540584\\
   0.785000000000000   0.733885014720019\\
   0.790000000000000   0.739611884144194\\
   0.795000000000000   0.745361975151130\\
   0.800000000000000   0.751135210136154\\
   0.805000000000000   0.756931356745290\\
   0.810000000000000   0.762750649075454\\
   0.815000000000000   0.768592843897473\\
   0.820000000000000   0.774457841310901\\
   0.825000000000000   0.780345589050343\\
   0.830000000000000   0.786255956624599\\
   0.835000000000000   0.792188879121285\\
   0.840000000000000   0.798144271645332\\
   0.845000000000000   0.804122043270236\\
   0.850000000000000   0.810122101298934\\
   0.855000000000000   0.816144351741714\\
   0.860000000000000   0.822188703577848\\
   0.865000000000000   0.828255064003797\\
   0.870000000000000   0.834343337314644\\
   0.875000000000000   0.840453428818569\\
   0.880000000000000   0.846585233204879\\
   0.885000000000000   0.852738661036627\\
   0.890000000000000   0.858913611955299\\
   0.895000000000000   0.865109987085826\\
   0.900000000000000   0.871327682265683\\
   0.905000000000000   0.877566599535815\\
   0.910000000000000   0.883826688853482\\
   0.915000000000000   0.890107726771095\\
   0.920000000000000   0.896409736896080\\
   0.925000000000000   0.902732526405676\\
   0.930000000000000   0.909076055924112\\
   0.935000000000000   0.915440126201910\\
   0.940000000000000   0.921824725945330\\
   0.945000000000000   0.928229698137722\\
   0.950000000000000   0.934654997664982\\
   0.955000000000000   0.941100450883782\\
   0.960000000000000   0.947565985504570\\
   0.965000000000000   0.954051486354846\\
   0.970000000000000   0.960556846128583\\
   0.975000000000000   0.967081840454730\\
   0.980000000000000   0.973626586808245\\
   0.985000000000000   0.980190931232022\\
   0.990000000000000   0.986774668949662\\
   0.995000000000000   0.993377639899889\\
   1.000000000000000   0.999999954053209\\
        }; \addlegendentry{Case 2}
      \end{axis}
    \end{tikzpicture}
\caption{Alice and Bob share the qutrit-qutrit state $\rho = p\ketbra{\phi^+}{\phi^+} + (1-p)\mathds{1}/9$. In case 1 the set of input states is $\{\ket{0},\ket{1},(\ket{0}+\ket{1}+\ket{2})/\sqrt{3},(\ket{0}+w\ket{1}+w^*\ket{2})/\sqrt{3}\}$, while in the case 2 the set of input states is $\{\ket{0},\ket{1},(\ket{0}+\ket{1}+\ket{2})/\sqrt{3},(\ket{0}+w\ket{1}+w^*\ket{2})/\sqrt{3},(\ket{0}+\ket{1}+w\ket{2})/\sqrt{3},(w\ket{0}+\ket{1}+\ket{2})/\sqrt{3}\}$, where $w = \exp{i2\pi/3}$. The graph shows the lower bounds derived from the knowledge of the whole set of teleported states on the self-tested fidelity with the maximally entangled pair of qutrits as a function of the parameter $p$. In none of two cases the set of input states is tomographically complete, hence no conclusion about the fidelity of the shared state with maximally entangled pair of qutrits can be drawn from the observed average teleportation fidelity.}
\label{fig:CompCHSH}
\end{figure}
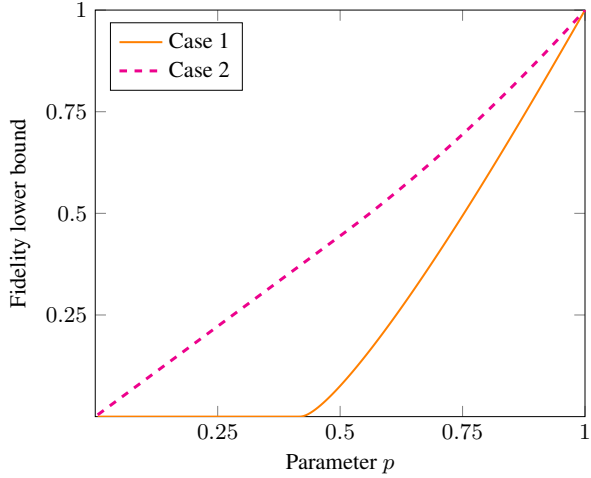

Thus, the lower bound on the fidelity between the physical state and $\ket{\phi^+}$ can be obtained as a solution to the following semi-definite programming (SDP) optimization:
\begin{align}\nonumber
    \min \qquad &\frac{1}{d}\sum_a\bra{\phi^+}U_a^{\rA'}{\tilde{\M}_a}^{T_{\rA'}}{U_a^{\dagger}}^{\rA'}\ket{\phi^+}\\ \label{sdp}
    \textrm{s.t} \qquad &\varphi_{a|x} = \textrm{Tr}_{\rA'}\left[\tilde{\M}_a^{\rA'\rB}(\psi_x^{\rA'}\otimes\mathds{1}^\rB)\right] \quad \forall a,x,\\ \nonumber
    &{\tilde{\M}_a}^{T_{\rA'}} \geq 0,\quad \forall a,  \qquad \sum_{a}\tilde{\M}_a = \mathds{1}\otimes\sum_a\varphi_{a|x},\quad \forall x.
\end{align}
The SDP \eqref{sdp} provides a lower bound on the fidelity between the physical state and $\ket{\phi^+}$ from the full observed data in a teleportation experiment. In principle the knowledge of the whole set of teleported states $\{\varphi_{a|x}\}_{a,x}$ is not necessary. One can fix some of the known teleportation quantifers, such as average teleportation fidelity, teleportation weight or one of the teleportation robustness measures introduced in \cite{isc2}. In Fig. \ref{fig:CompCHSH} we solve the SDP in \eqref{sdp} for two cases without a tomographically complete set of states, two situations where the average teleportation fidelity cannot be used.

\subsection{Self-testing of quantum networks}\label{networkst}

\begin{figure*}
\centering
\includegraphics[width=2\columnwidth]{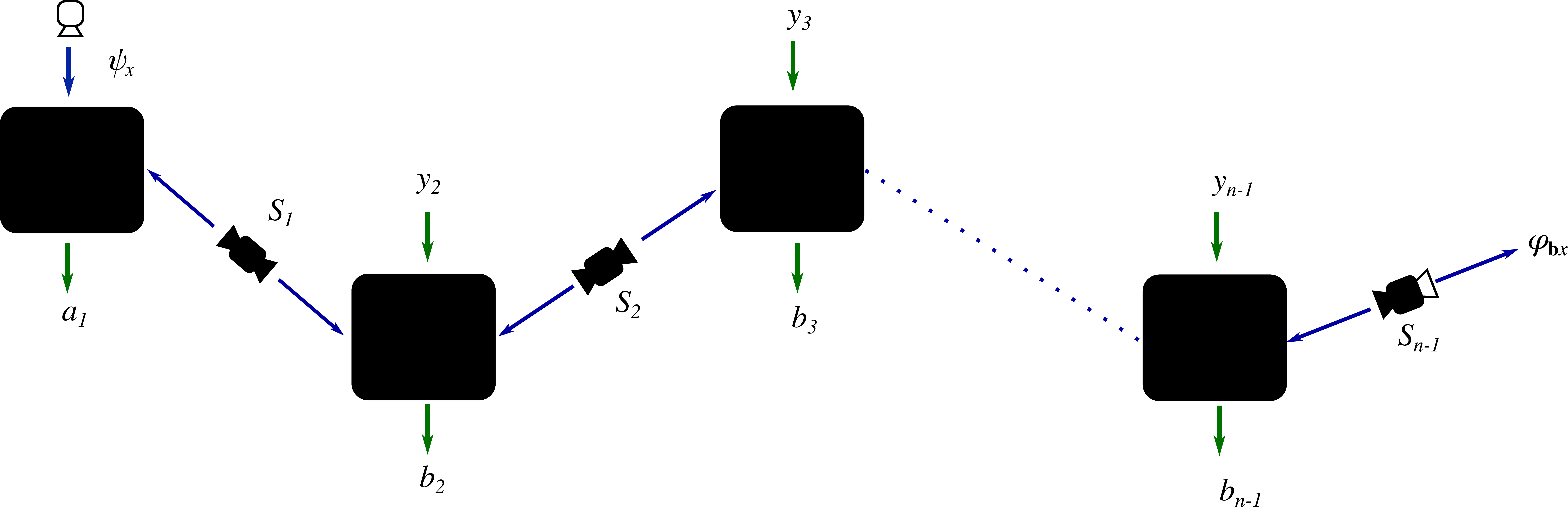}
\caption{A networked scenario where trusted quantum systems can be input into untrusted devices at the beginning and trusted quantum systems can be measured at the end. Intermediate, untrusted nodes can be used to teleport a state, or use quantum repeaters to establish entanglement. Techniques developed here can be used to certify the whole network along with individual links. \label{fig:network}}
\end{figure*}

Equipped with the methods presented in the previous sections, we are in position to provide ways of self-testing elements of a quantum network. Complementary to the results about self-testing Bell state measurements~\cite{mo,jd}, we provide means to self-test different links of potentially hybrid  quantum network. Consider a network in the form of a quantum repeater, like the one on Fig. \ref{fig:network}. All measurement devices, except the first and the last, have a classical input whose choice corresponds to a Bell state measurement on the two particles or measuring one of the particles shared with one of the neighbours. One might extend our method for self-testing from teleportation and find out how well the whole quantum repeater simulates a single maximally entangled state.  If the fidelity is not satisfactory, it is possible to check separate links of the network. For example, the `quality' of the source $S_1$  can be estimated by using the self-testing with quantum-classical inputs (section \ref{qcst}). Self-testing through EPR steering \cite{IvanMatty} can be used to self-test  source $S_{n-1}$. Standard self-testing protocols can be used to self-test all the remaining sources.

\section*{Discussion}
In this work we have expanded quantum state certification to novel scenarios using quantum inputs. Developing a hybrid approach between full device-independent  and device-dependent self-testing is one of the main motivations of this work, with applications to quantum networks where some nodes in the network are trusted, and others are not. The tools developed here in the MDI setting could also find an application in a networked device-independent setting using the ideas developed in~\cite{bowles}. 

This approach also finds an application of recent work in the study of non-classical teleportation introduced by \cite{tel}. In particular, we have developed new numerical tools to relate quantum teleportation to the fidelity of the quantum states shared by the parties. Given the ubiquity of teleportation in quantum information processing, these tools could be used in the verification of teleportation-based quantum computing. 

One direction for future research is exploring the set of quantum correlations in different scenarios with quantum inputs. This would open the doors for numerical self-testing, similar to the SWAP method from \cite{swap1,swap2} or the numerical self-test presented in our Section \ref{tlpst}. Another interesting question is to search better isometries for self-testing than those considered in this work. Finding a good isometry is crucial for obtaining better noise-resistant self-testing protocols. In turn, this could make self-testing more applicable and practical.

\textit{Note}--- While finishing this manuscript we became aware of a similar work \cite{zhang2019simultaneous}.

\section*{Acknowledgements}
This work was supported by  SNSF (Bridge project 'Self-testing QRNG'), COST project CA16218, NANOCOHYBRI.  Spanish  MINECO  (QIBEQI FIS2016-80773-P  and  Severo  Ochoa  SEV-2015-0522),  the  AXA  Chair  in  Quantum
Information  Science,   Generalitat  de  Catalunya  (CERCA  Programme),   Fundacio
Privada Cellex and ERC CoG QITBOX. MJH acknowledges the FQXi large grant The Emergence of Agents from Causal Order.

\bibliographystyle{alphaurl}
\bibliography{sample}

\newcommand{\etalchar}[1]{$^{#1}$}
\begin{thebibliography}{BCMdW10}

\bibitem[AB09]{andersbrowne}
Janet Anders and Dan~E. Browne.
\newblock Computational power of correlations.
\newblock {\em Phys. Rev. Lett.}, 102:050502, Feb 2009.
\newblock \href {http://dx.doi.org/10.1103/PhysRevLett.102.050502}
  {\path{doi:10.1103/PhysRevLett.102.050502}}.

\bibitem[ABG{\etalchar{+}}07]{diqkd}
Antonio Ac{\'\i}n, Nicolas Brunner, Nicolas Gisin, Serge Massar, Stefano
  Pironio, and Valerio Scarani.
\newblock Device-independent security of quantum cryptography against
  collective attacks.
\newblock {\em Physical Review Letters}, 98(23):230501, 2007.
\newblock \href {http://dx.doi.org/10.1103/PhysRevLett.98.230501}
  {\path{doi:10.1103/PhysRevLett.98.230501}}.

\bibitem[AM16]{randomnessreview}
Antonio Ac{\'\i}n and Lluis Masanes.
\newblock Certified randomness in quantum physics.
\newblock {\em Nature}, 540:213--219, 2016.
\newblock \href {http://dx.doi.org/10.1038/nature20119}
  {\path{doi:10.1038/nature20119}}.

\bibitem[BBC{\etalchar{+}}93]{teleportation}
Charles~H. Bennett, Gilles Brassard, Claude Cr\'epeau, Richard Jozsa, Asher
  Peres, and William~K. Wootters.
\newblock Teleporting an unknown quantum state via dual classical and
  einstein-podolsky-rosen channels.
\newblock {\em Physical Review Letters}, 70:1895--1899, Mar 1993.
\newblock \href {http://dx.doi.org/10.1103/PhysRevLett.70.1895}
  {\path{doi:10.1103/PhysRevLett.70.1895}}.

\bibitem[BCMdW10]{communication}
Harry Buhrman, Richard Cleve, Serge Massar, and Ronald de~Wolf.
\newblock Nonlocality and communication complexity.
\newblock {\em Rev. Mod. Phys.}, 82:665--698, Mar 2010.
\newblock \href {http://dx.doi.org/10.1103/RevModPhys.82.665}
  {\path{doi:10.1103/RevModPhys.82.665}}.

\bibitem[BCP{\etalchar{+}}14]{review}
Nicolas Brunner, Daniel Cavalcanti, Stefano Pironio, Valerio Scarani, and
  Stephanie Wehner.
\newblock Bell nonlocality.
\newblock {\em Rev. Mod. Phys.}, 86:419--478, Apr 2014.
\newblock \href {http://dx.doi.org/10.1103/RevModPhys.86.419}
  {\path{doi:10.1103/RevModPhys.86.419}}.

\bibitem[BDCZ98]{repeaters}
H.-J. Briegel, W.~D\"ur, J.~I. Cirac, and P.~Zoller.
\newblock Quantum repeaters: The role of imperfect local operations in quantum
  communication.
\newblock {\em Phys. Rev. Lett.}, 81:5932--5935, Dec 1998.
\newblock \href {http://dx.doi.org/10.1103/PhysRevLett.81.5932}
  {\path{doi:10.1103/PhysRevLett.81.5932}}.

\bibitem[Bel64]{bell}
John~Stewart Bell.
\newblock On the {E}instein-{P}odolsky-{R}osen paradox.
\newblock {\em Physics}, 1:195--200, 1964.
\newblock URL: \url{https://cds.cern.ch/record/111654}.

\bibitem[BHK05]{bhk}
Jonathan Barrett, Lucien Hardy, and Adrian Kent.
\newblock No signaling and quantum key distribution.
\newblock {\em Phys. Rev. Lett.}, 95:010503, Jun 2005.
\newblock \href {http://dx.doi.org/10.1103/PhysRevLett.95.010503}
  {\path{doi:10.1103/PhysRevLett.95.010503}}.

\bibitem[BNS{\etalchar{+}}15]{swap2}
Jean-Daniel Bancal, Miguel Navascu\'es, Valerio Scarani, Tam\'as V\'ertesi, and
  Tzyh~Haur Yang.
\newblock Physical characterization of quantum devices from nonlocal
  correlations.
\newblock {\em Phys. Rev. A}, 91:022115, Feb 2015.
\newblock \href {http://dx.doi.org/10.1103/PhysRevA.91.022115}
  {\path{doi:10.1103/PhysRevA.91.022115}}.

\bibitem[BRLG13]{mdi}
Cyril Branciard, Denis Rosset, Yeong-Cherng Liang, and Nicolas Gisin.
\newblock Measurement-device-independent entanglement witnesses for all
  entangled quantum states.
\newblock {\em Phys. Rev. Lett.}, 110:060405, Feb 2013.
\newblock \href {http://dx.doi.org/10.1103/PhysRevLett.110.060405}
  {\path{doi:10.1103/PhysRevLett.110.060405}}.

\bibitem[B{\v{S}}CA18]{bowles}
Joseph Bowles, Ivan {\v{S}}upi\'{c}, Daniel Cavalcanti, and Antonio Ac\'{\i}n.
\newblock Device-independent entanglement certification of all entangled
  states.
\newblock {\em Phys. Rev. Lett.}, 121:180503, Oct 2018.
\newblock \href {http://dx.doi.org/10.1103/PhysRevLett.121.180503}
  {\path{doi:10.1103/PhysRevLett.121.180503}}.

\bibitem[BSS18]{jd}
Jean-Daniel Bancal, Nicolas Sangouard, and Pavel Sekatski.
\newblock Noise-resistant device-independent certification of {B}ell state
  measurements.
\newblock {\em Phys. Rev. Lett.}, 121:250506, Dec 2018.
\newblock \href {http://dx.doi.org/10.1103/PhysRevLett.121.250506}
  {\path{doi:10.1103/PhysRevLett.121.250506}}.

\bibitem[Bus12]{buscemi}
Francesco Buscemi.
\newblock All entangled quantum states are nonlocal.
\newblock {\em Phys. Rev. Lett.}, 108:200401, May 2012.
\newblock \href {http://dx.doi.org/10.1103/PhysRevLett.108.200401}
  {\path{doi:10.1103/PhysRevLett.108.200401}}.

\bibitem[CGS17]{Coladangelo2017}
Andrea Coladangelo, Koon~Tong Goh, and Valerio Scarani.
\newblock All pure bipartite entangled states can be self-tested.
\newblock {\em Nature Communications}, 8:15485, may 2017.
\newblock \href {http://dx.doi.org/10.1038/ncomms15485}
  {\path{doi:10.1038/ncomms15485}}.

\bibitem[Col06]{colbeck}
Roger Colbeck.
\newblock {\em Quantum And Relativistic Protocols For Secure Multi-Party
  Computation}.
\newblock PhD thesis, University of Cambridge, 2006.
\newblock URL: \url{https://arxiv.org/abs/0911.3814}.

\bibitem[CS17]{steeringreview}
Daniel Cavalcanti and Paul Skrzypczyk.
\newblock Quantum steering: a review with focus on semidefinite programming.
\newblock {\em Reports on Progress in Physics}, 80(2):024001, 2017.
\newblock \href {http://dx.doi.org/10.1088/1361-6633/80/2/024001}
  {\path{doi:10.1088/1361-6633/80/2/024001}}.

\bibitem[CS{\v{S}}17]{tel}
Daniel Cavalcanti, Paul Skrzypczyk, and Ivan {\v{S}}upi{\'{c}}.
\newblock All entangled states can demonstrate nonclassical teleportation.
\newblock {\em Phys. Rev. Lett.}, 119:110501, Sep 2017.
\newblock \href {http://dx.doi.org/10.1103/PhysRevLett.119.110501}
  {\path{doi:10.1103/PhysRevLett.119.110501}}.

\bibitem[Eke91]{ekert}
Artur~K. Ekert.
\newblock Quantum cryptography based on {B}ell's theorem.
\newblock {\em Physical Review Letters}, 67:661--663, Aug 1991.
\newblock \href {http://dx.doi.org/10.1103/PhysRevLett.67.661}
  {\path{doi:10.1103/PhysRevLett.67.661}}.

\bibitem[FK19]{Farkas}
M\'at\'e Farkas and J\ifmmode\mbox{\k{e}}\else~\k{e}\fi{}drzej Kaniewski.
\newblock Self-testing mutually unbiased bases in the prepare-and-measure
  scenario.
\newblock {\em Phys. Rev. A}, 99:032316, Mar 2019.
\newblock \href {http://dx.doi.org/10.1103/PhysRevA.99.032316}
  {\path{doi:10.1103/PhysRevA.99.032316}}.

\bibitem[GBHA10]{Gallego}
Rodrigo Gallego, Nicolas Brunner, Christopher Hadley, and Antonio Ac\'{\i}n.
\newblock Device-independent tests of classical and quantum dimensions.
\newblock {\em Phys. Rev. Lett.}, 105:230501, Nov 2010.
\newblock \href {http://dx.doi.org/10.1103/PhysRevLett.105.230501}
  {\path{doi:10.1103/PhysRevLett.105.230501}}.

\bibitem[GKW15]{Gheorghiu}
Alexandru Gheorghiu, Elham Kashefi, and Petros Wallden.
\newblock Robustness and device independence of verifiable blind quantum
  computing.
\newblock {\em New Journal of Physics}, 17(8):083040, 2015.
\newblock \href {http://dx.doi.org/10.1088/1367-2630/17/8/083040}
  {\path{doi:10.1088/1367-2630/17/8/083040}}.

\bibitem[GKW{\etalchar{+}}18]{Goh}
Koon~Tong Goh, Jędrzej Kaniewski, Elie Wolfe, Tam\'as V\'ertesi, Xingyao Wu,
  Yu~Cai, Yeong-Cherng Liang, and Valerio Scarani.
\newblock Geometry of the set of quantum correlations.
\newblock {\em Phys. Rev. A}, 97:022104, Feb 2018.
\newblock \href {http://dx.doi.org/10.1103/PhysRevA.97.022104}
  {\path{doi:10.1103/PhysRevA.97.022104}}.

\bibitem[GWK17]{Gheorghiu1}
Alexandru Gheorghiu, Petros Wallden, and Elham Kashefi.
\newblock Rigidity of quantum steering and one-sided device-independent
  verifiable quantum computation.
\newblock {\em New Journal of Physics}, 19(2):023043, feb 2017.
\newblock \href {http://dx.doi.org/10.1088/1367-2630/aa5cff}
  {\path{doi:10.1088/1367-2630/aa5cff}}.

\bibitem[HB11]{hoban}
Matty~J Hoban and Dan~E Browne.
\newblock Stronger quantum correlations with loophole-free postselection.
\newblock {\em Phys. Rev. Lett.}, 107(12):120402, Sep 2011.
\newblock \href {http://dx.doi.org/10.1103/PhysRevLett.107.120402}
  {\path{doi:10.1103/PhysRevLett.107.120402}}.

\bibitem[HHH99]{HHH}
Micha\l{} Horodecki, Pawe\l{} Horodecki, and Ryszard Horodecki.
\newblock General teleportation channel, singlet fraction, and
  quasidistillation.
\newblock {\em Phys. Rev. A}, 60:1888--1898, Sep 1999.
\newblock \href {http://dx.doi.org/10.1103/PhysRevA.60.1888}
  {\path{doi:10.1103/PhysRevA.60.1888}}.

\bibitem[HHHH09]{entanglementreview}
Ryszard Horodecki, Pawe\l{} Horodecki, Micha\l{} Horodecki, and Karol
  Horodecki.
\newblock Quantum entanglement.
\newblock {\em Review of Modern Physics}, 81:865--942, Jun 2009.
\newblock \href {http://dx.doi.org/10.1103/RevModPhys.81.865}
  {\path{doi:10.1103/RevModPhys.81.865}}.

\bibitem[HS18]{mattybelen}
Matty~J Hoban and Ana~Bel{\'{e}}n Sainz.
\newblock A channel-based framework for steering, non-locality and beyond.
\newblock {\em New Journal of Physics}, 20(5):053048, 2018.
\newblock \href {http://dx.doi.org/10.1088/1367-2630/aabea8}
  {\path{doi:10.1088/1367-2630/aabea8}}.

\bibitem[Kan17]{Jed1}
Jędrzej Kaniewski.
\newblock Self-testing of binary observables based on commutation.
\newblock {\em Phys. Rev. A}, 95:062323, Jun 2017.
\newblock \href {http://dx.doi.org/10.1103/PhysRevA.95.062323}
  {\path{doi:10.1103/PhysRevA.95.062323}}.

\bibitem[McK14]{McKague2014}
Mathew McKague.
\newblock Self-testing graph states.
\newblock In D.~Bacon, M.~Martin-Delgado, and M.~Roetteler, editors, {\em
  Theory of Quantum Computation, Communication, and Cryptography ,}, volume
  6745 of {\em Lecture Notes in Computer Science}, pages 104--120. Springer,
  Berlin, Heidelberg, 2014.
\newblock \href {http://dx.doi.org/https://doi.org/10.1007/978-3-642-54429-3_7}
  {\path{doi:https://doi.org/10.1007/978-3-642-54429-3_7}}.

\bibitem[MK18]{markham}
Damian Markham and Alexandra Krause.
\newblock {A} simple protocol for certifying graph states and applications in
  quantum networks, 2018.
\newblock \href {http://arxiv.org/abs/1801.05057} {\path{arXiv:1801.05057}}.

\bibitem[MY04]{Mayers2004}
D.~Mayers and A.~Yao.
\newblock Self testing quantum apparatus.
\newblock {\em Quantum Info. Comput.}, 4:273, 2004.

\bibitem[PAM{\etalchar{+}}10]{pironio2010random}
Stefano Pironio, Antonio Ac{\'\i}n, Serge Massar, A~Boyer de~La~Giroday,
  Dzimitry~N Matsukevich, Peter Maunz, Steven Olmschenk, David Hayes, Le~Luo,
  T~Andrew Manning, et~al.
\newblock Random numbers certified by {B}ell’s theorem.
\newblock {\em Nature}, 464(7291):1021--1024, 2010.
\newblock \href {http://dx.doi.org/10.1038/nature09008}
  {\path{doi:10.1038/nature09008}}.

\bibitem[PB11]{BP}
Marcin Paw\l{}owski and Nicolas Brunner.
\newblock Semi-device-independent security of one-way quantum key distribution.
\newblock {\em Phys. Rev. A}, 84:010302, Jul 2011.
\newblock \href {http://dx.doi.org/10.1103/PhysRevA.84.010302}
  {\path{doi:10.1103/PhysRevA.84.010302}}.

\bibitem[PLM18]{ashley}
Sam Pallister, Noah Linden, and Ashley Montanaro.
\newblock Optimal verification of entangled states with local measurements.
\newblock {\em Phys. Rev. Lett.}, 120:170502, Apr 2018.
\newblock \href {http://dx.doi.org/10.1103/PhysRevLett.120.170502}
  {\path{doi:10.1103/PhysRevLett.120.170502}}.

\bibitem[PR04]{qse}
Matteo Paris and Jaroslav Rehacek.
\newblock {\em Quantum State Estimation}, volume 649 of {\em Lecture Notes in
  Physics}.
\newblock Springer-Verlag Berlin Heidelberg, 2004.
\newblock \href {http://dx.doi.org/10.1007/b98673} {\path{doi:10.1007/b98673}}.

\bibitem[Pru77]{Prugovecki1977}
E.~Prugove{\v{c}}ki.
\newblock Information-theoretical aspects of quantum measurement.
\newblock {\em International Journal of Theoretical Physics}, 16(5):321--331,
  May 1977.
\newblock \href {http://dx.doi.org/10.1007/BF01807146}
  {\path{doi:10.1007/BF01807146}}.

\bibitem[RKB18]{mo}
Marc~Olivier Renou, J\ifmmode \mbox{\k{e}}\else~\k{e}\fi{}drzej Kaniewski, and
  Nicolas Brunner.
\newblock Self-testing entangled measurements in quantum networks.
\newblock {\em Phys. Rev. Lett.}, 121:250507, Dec 2018.
\newblock \href {http://dx.doi.org/10.1103/PhysRevLett.121.250507}
  {\path{doi:10.1103/PhysRevLett.121.250507}}.

\bibitem[RMV{\etalchar{+}}18]{RMVLT}
Denis Rosset, Anthony Martin, Ephanielle Verbanis, Charles Ci~Wen Lim, and Rob
  Thew.
\newblock Practical measurement-device-independent entanglement quantification.
\newblock {\em Phys. Rev. A}, 98:052332, Nov 2018.
\newblock \href {http://dx.doi.org/10.1103/PhysRevA.98.052332}
  {\path{doi:10.1103/PhysRevA.98.052332}}.

\bibitem[RUV13]{ruv}
Ben~W. Reichardt, Falk Unger, and Umesh Vazirani.
\newblock Classical command of quantum systems.
\newblock {\em Nature}, 496:456, 2013.
\newblock \href {http://dx.doi.org/10.1038/nature12035}
  {\path{doi:10.1038/nature12035}}.

\bibitem[{\v{S}}B19]{STreview}
Ivan {\v{S}}upi{\'{c}} and Joseph Bowles.
\newblock Self-testing of quantum systems: a review, 2019.
\newblock \href {http://arxiv.org/abs/1904.10042} {\path{arXiv:1904.10042}}.

\bibitem[{\v{S}}CAA18]{Ivan}
Ivan {\v{S}}upi{\'{c}}, Andrea Coladangelo, Remigiusz Augusiak, and Antonio
  Acín.
\newblock Self-testing multipartite entangled states through projections onto
  two systems.
\newblock {\em New Journal of Physics}, 20(8):083041, 2018.
\newblock \href {http://dx.doi.org/10.1088/1367-2630/aad89b}
  {\path{doi:10.1088/1367-2630/aad89b}}.

\bibitem[SCB{\etalchar{+}}19]{Weixu}
Weixu Shi, Yu~Cai, Jonatan~Bohr Brask, Hugo Zbinden, and Nicolas Brunner.
\newblock {S}emi-device-independent characterization of quantum measurements
  under a minimum overlap assumption, 2019.
\newblock \href {http://arxiv.org/abs/1904.05692} {\path{arXiv:1904.05692}}.

\bibitem[Sch35]{erwin}
E.~Schrödinger.
\newblock Discussion of probability relations between separated systems.
\newblock {\em Mathematical Proceedings of the Cambridge Philosophical
  Society}, 31(4):555–563, 1935.
\newblock \href {http://dx.doi.org/10.1017/S0305004100013554}
  {\path{doi:10.1017/S0305004100013554}}.

\bibitem[{\v{S}}H16]{IvanMatty}
Ivan {\v{S}}upi\'{c} and Matty~J Hoban.
\newblock Self-testing through epr-steering.
\newblock {\em New Journal of Physics}, 18(7):075006, 2016.
\newblock \href {http://dx.doi.org/10.1088/1367-2630/18/7/075006}
  {\path{doi:10.1088/1367-2630/18/7/075006}}.

\bibitem[{\v{S}}SC17]{IPD}
Ivan {\v{S}}upi{\'{c}}, Paul Skrzypczyk, and Daniel Cavalcanti.
\newblock Measurement-device-independent entanglement and randomness estimation
  in quantum networks.
\newblock {\em Phys. Rev. A}, 95:042340, Apr 2017.
\newblock \href {http://dx.doi.org/10.1103/PhysRevA.95.042340}
  {\path{doi:10.1103/PhysRevA.95.042340}}.

\bibitem[{\v{S}}SC19]{isc2}
Ivan {\v{S}}upi\'{c}, Paul Skrzypczyk, and Daniel Cavalcanti.
\newblock Methods to estimate entanglement in teleportation experiments.
\newblock {\em Phys. Rev. A}, 99:032334, Mar 2019.
\newblock \href {http://dx.doi.org/10.1103/PhysRevA.99.032334}
  {\path{doi:10.1103/PhysRevA.99.032334}}.

\bibitem[SVW16]{SVZ}
Jamie Sikora, Antonios Varvitsiotis, and Zhaohui Wei.
\newblock Minimum dimension of a hilbert space needed to generate a quantum
  correlation.
\newblock {\em Phys. Rev. Lett.}, 117:060401, Aug 2016.
\newblock \href {http://dx.doi.org/10.1103/PhysRevLett.117.060401}
  {\path{doi:10.1103/PhysRevLett.117.060401}}.

\bibitem[TKV{\etalchar{+}}18]{Armin}
Armin Tavakoli, J\ifmmode \mbox{\k{e}}\else~\k{e}\fi{}drzej Kaniewski, Tam\'as
  V\'ertesi, Denis Rosset, and Nicolas Brunner.
\newblock Self-testing quantum states and measurements in the
  prepare-and-measure scenario.
\newblock {\em Phys. Rev. A}, 98:062307, Dec 2018.
\newblock \href {http://dx.doi.org/10.1103/PhysRevA.98.062307}
  {\path{doi:10.1103/PhysRevA.98.062307}}.

\bibitem[TM18]{TakeuchiMorimae}
Yuki Takeuchi and Tomoyuki Morimae.
\newblock Verification of many-qubit states.
\newblock {\em Phys. Rev. X}, 8:021060, Jun 2018.
\newblock \href {http://dx.doi.org/10.1103/PhysRevX.8.021060}
  {\path{doi:10.1103/PhysRevX.8.021060}}.

\bibitem[UCNG19]{steeringreview2}
Roope Uola, Ana C.~S. Costa, H.~Chau Nguyen, and Otfried Gühne.
\newblock {Q}uantum {S}teering, 2019.
\newblock \href {http://arxiv.org/abs/1903.06663} {\path{arXiv:1903.06663}}.

\bibitem[WJD07]{wiseman}
H.~M. Wiseman, S.~J. Jones, and A.~C. Doherty.
\newblock Steering, entanglement, nonlocality, and the einstein-podolsky-rosen
  paradox.
\newblock {\em Phys. Rev. Lett.}, 98:140402, Apr 2007.
\newblock \href {http://dx.doi.org/10.1103/PhysRevLett.98.140402}
  {\path{doi:10.1103/PhysRevLett.98.140402}}.

\bibitem[YVB{\etalchar{+}}14]{swap1}
Tzyh~Haur Yang, Tam\'as V\'ertesi, Jean-Daniel Bancal, Valerio Scarani, and
  Miguel Navascu\'es.
\newblock Robust and versatile black-box certification of quantum devices.
\newblock {\em Phys. Rev. Lett.}, 113:040401, Jul 2014.
\newblock \href {http://dx.doi.org/10.1103/PhysRevLett.113.040401}
  {\path{doi:10.1103/PhysRevLett.113.040401}}.

\bibitem[ZZ19]{zhang2019simultaneous}
Xingjian Zhang and Qi~Zhao.
\newblock Simultaneous certification of entangled states and measurements in
  bounded dimensional semi-quantum games, 2019.
\newblock \href {http://arxiv.org/abs/1911.05981} {\path{arXiv:1911.05981}}.

\end{thebibliography}

\newpage

\pagebreak
\widetext
\begin{appendix}
\setcounter{equation}{0}
\setcounter{figure}{0}
\setcounter{table}{0}
\setcounter{page}{1}
\setcounter{section}{0}
\makeatletter
\renewcommand{\theequation}{S\arabic{equation}}
\renewcommand{\thefigure}{S\arabic{figure}}
\renewcommand{\bibnumfmt}[1]{[S#1]}
\renewcommand{\citenumfont}[1]{S#1}

\section{A remark about mixed states}\label{mixed}

Here we provide a proof that the correlations with quantum inputs obtained  by measuring any mixed state can be obtained by using a pure state of the same local dimensions. It is analogous to the proof from \cite{SVZ}.
\begin{proof}
Suppose that $\{p(a,b|\psi_x,\psi_y) = \textrm{Tr}((\M_a^{\rA'\rA} \otimes \M_b^{\rB\rB'})\cdot \psi_x\otimes {\varrho'}^{\rA\rB} \otimes \psi_y)\}$ are generated by a mixed state ${\varrho'}^{\rA\rB}$ acting on $\mathcal{H}_\rA \otimes \mathcal{H}_\rB$. Without loss of generality we consider the case $d_A\leq d_B$. Let $\ket{\tilde{\phi}} \in \mathcal{H}_\rA \otimes \mathcal{H}_\rB \otimes \mathcal{H}_\mathsf{P}$ be the purification of ${\varrho'}^{\rA\rB}$, and let $\ket{\tilde{\phi}} = \sum_{i=1}^{d_A} \lambda_i \ket{\mathrm{a}^i_\rA}\ket{\mathrm{b}^i_{\rB\mathsf{P}}}$ be its Schmidt decomposition, where $\ket{\mathrm{a}^i_\rA} \in \mathcal{H}_\rA$ and $\ket{\mathrm{b}^i_{\rB\mathsf{P}}} \in \mathcal{H}_\rB \otimes \mathcal{H}_\mathsf{P}$.  Define the isometry  $V = \sum_{i=1}^{d_\rA} \ketbra{i_{\rB}}{\mathrm{b}^i_{\rB\mathsf{P}}}$ from $\mathcal{H}_\rB \otimes \mathcal{H}_\mathsf{P}$ to $\mathcal{H}_\rB$. Note that $\ket{\phi} =(\mathds{1} \otimes V)\ket{\tilde{\phi}} = \sum_{i=1}^{d_\rA}\lambda_i \ket{\mathrm{a}^i_\rA}\ket{i_{\rB}}\in\mathcal{H}_\rA \otimes \mathcal{H}_\rB$.  One can see that the operators $\{\bar{\M}_b^{\rB\rB'} \otimes \mathds{1}^\mathsf{P} = (V \otimes \mathds{1}^{\rB'})(\M_b^{\rB\rB'}\otimes \mathds{1}^{\mathsf{P}})(V^\dag \otimes \mathds{1}^{\rB'})\}$ define a projective measurement acting on  $\rB\rB'$. Finally,
\[
\textrm{Tr}((\M_a^{\rA'\rA} \otimes \M_b^{\rB\rB'})(\psi_x^{\rA'}\otimes {\varrho'}^{\rA\rB} \otimes \psi_y^{\rB'}) =\bra{\psi_x} \otimes \bra{\phi} \otimes \bra{\psi_y}({\M}_a^{\rA'\rA} \otimes \bar{\M}_b^{\rB\rB'})\ket{\psi_x} \otimes \ket{\phi} \otimes \ket{\psi_y} .
\]
\end{proof}

\section{Proof of Theorem 1}\label{proof1}

For the sake of convenience, we repeat Theorem 1 here.\\
\textbf{Theorem 1} $\quad$
\textit{Let two parties, Alice and Bob, share  the state ${\varrho'}^\emph{\rA\rB}$ and have access to a tomographically complete set of inputs $\{\psi_x\}_x$ and $\{\psi_y\}_y$ respectively. Each party performs a joint measurement on their share of ${\varrho'}^\emph{\rA\rB}$ and quantum input $\psi_{x}$ or $\psi_{y}$. If the correlation probabilities can be written in the form
\begin{eqnarray}\label{1sm}
p\left(a,b \vert \psi_x,\psi_y\right) = \emph{Tr}\left[{\tilde{\M}_{a,b}}^{\emph{\rA}'\emph{\rB}'}\left(\psi_x^{\emph{\rA}'}\otimes \psi_y^{\emph{\rB}'}\right)\right],  \quad \forall a,b,x,y;
\end{eqnarray}
and $\tilde{\M}^{\emph{\rA}'\emph{\rB}'}_{a,b}$ are such that
\begin{eqnarray}\label{2sm}
\frac{1}{d^2}\ketbra{\psi}{\psi} &=& (U_a\otimes U_b){(\tilde{\M}_{a,b}^{\emph{\rA}'\emph{\rB}'})}^T(U_a^{\dagger}\otimes U_b^{\dagger})  \quad \forall a,b,
\end{eqnarray}
where $U_a$ and $U_b$ are the correcting unitaries defined as $U_m = \sum_{kl}X^kZ^l\delta_{m,kl}$, then there exists a local isometry $\Phi$ such that 
\begin{equation}\label{statementsm}
\Phi({{\varrho'}^\emph{\rA\rB}}) = \ketbra{\psi}{\psi}^{\emph{\rA}''\emph{\rB}''}\otimes \varrho_{junk}^{\emph{\rA\rA}'\emph{\rB\rB}'}
\end{equation}}
\qed \\

Since the set of quantum inputs is tomographically complete Eqs. (\ref{1sm}) imply that all effective measurements defined as
\begin{equation*}
\tilde{\M}_{a,b} = \textrm{Tr}_{\rA\rB}\left[\left(\M_a^{\rA'\rA}\otimes \M_b^{\rB\rB'}\right)\left(\mathds{1}^{\rA'}\otimes \varrho^{\rA\rB}\otimes \mathds{1}^{\rB'}\right)\right]
\end{equation*}
are proportional to rank-one projective operators satisfying constraints (\ref{2sm}). Consider the isometry shown in Figure \ref{iso} (in the main text). Applying $\Phi$ to ${\varrho'}^{\rA\rB}$ leads to
\begin{equation}
$$
\begin{multline*}
\Phi({\varrho'}^{\rA\rB}) = \Bigg(\sum_{a,b} U_a^{\rA''} \otimes \M_a^{\rA'\rA} \otimes \M_b^{\rB\rB'} \otimes U_b^{\rB''}\Bigg) \cdot \left(C_{\Sigma_x}^{\rA''\rA'} \cdot \left(F^{\rA''} \otimes \mathds{1}^{\rA'}\right) \otimes C_{\Sigma_x}^{\rB''\rB'}\cdot\left(\mathds{1}^{\rB'} \otimes F^{\rB''}  \right)\right) \cdot \bigg(\ketbra{00}{00}^{\rA'\rA''} \otimes \\ 
\otimes {\varrho'}^{\rA\rB} \otimes \ketbra{00}{00}^{\rB'\rB''} \bigg)\cdot \left(\left({F^{\dagger}}^{\rA''} \otimes \mathds{1}^{\rA'}\right)\cdot {C_{\Sigma_x}^{\dagger}}^{\rA''\rA'}  \otimes \left(\mathds{1}^{\rB'} \otimes {F^{\dagger}}^{\rB''}  \right)\cdot {C_{\Sigma_x}^{\dagger}}^{\rB''\rB'}\right) \cdot\Bigg(\sum_{a',b'} U_{a'}^{\dag \rA''} \otimes \M_{a'}^{\dag \rA'\rA} \otimes \M_{b'}^{\dag \rB\rB'} \otimes U_{b'}^{\dag \rB''}\Bigg)
\end{multline*}
\begin{multline*}
= \sum_{a,b,a',b'} \left(U_a^{\rA''} \otimes \M_a^{\rA'\rA} \otimes \M_b^{\rB\rB'} \otimes U_b^{\rB''}\right) \cdot \left({\phi^+}^{\rA'\rA''}\otimes {\varrho'}^{\rA\rB} \otimes {\phi^+}^{\rB'\rB''} \right)
\cdot \left(U_{a'}^{\dag \rA''} \otimes \M_{a'}^{\dag \rA'\rA} \otimes \M_{b'}^{\dag \rB'\rB} \otimes U_{b'}^{\dag \rB''}\right),
\end{multline*}
$$
\label{iso_exp}
\end{equation}
where $C_{\Sigma_x} = \sum_{j,k = 0}^{d-1}\ketbra{j}{j}\otimes\ketbra{k+j\mod d}{k}$ is the generlised CNOT gate, $\phi^+ = \ketbra{\phi^+}{\phi^+}$, and $\ket{\phi^+} = \frac{1}{\sqrt{d}}\sum_{j=0}^{d-1}\ket{jj}$ is the maximally entangled state. Now we can trace out $\rA\rA'\rB\rB'$ and see if the resulting state is pure. 

\begin{multline*}
\textrm{Tr}_{\rA\rA'\rB\rB'} \left[ \sum_{a,b,a',b'} \left(U_a^{\rA''} \otimes \M_a^{\rA'\rA} \otimes \M_b^{\rB\rB'} \otimes U_b^{\rB''}\right)\left({\phi^+}^{\rA'\rA''}\otimes {\varrho'}^{\rA\rB} \otimes {\phi^+}^{\rB'\rB''} \right) \left(U_{a'}^{\dag \rA''} \otimes \M_{a'}^{\dag \rA'\rA} \otimes \M_{b'}^{\dag \rB\rB'} \otimes U_{b'}^{\dag \rB''}\right)\right]
\end{multline*}
\begin{multline*}
= \sum_{a,b,a',b'} \left(U_a^{\rA''} \otimes U_b^{\rB''}\right) \tr_{\rA'\rA\rB\rB'}\Bigg[\left(\mathds{1}^{\rA''}\otimes \M_a^{\rA'\rA} \otimes \M_b^{\rB\rB'} \otimes \mathds{1}^{\rB''}\right) \left({\phi^+}^{\rA''\rA'} \otimes {\varrho'}^{\rA\rB} \otimes {\phi^+}^{\rB'\rB''}\right)\\
\left(\mathds{1}^{\rA''} \otimes \M_{a'}^{\dag \rA'\rA} \otimes \M_{b'}^{\dag \rB\rB'} \otimes \mathds{1}^{\rB''}\right)\Bigg]
  \left(U_{a'}^{\dag \rA''} \otimes U_{b'}^{\dag \rB''}\right)
\end{multline*}
\begin{eqnarray} \nonumber
&=& \sum_{a,b,a',b'} (U_a^{\rA''} \otimes U_b^{\rB''}) \tr_{\rA'\rA\rB\rB'}\left[\left(\mathds{1}^{\rA''} \otimes \M_{a'}^{\dag \rA\rA'} \M_a^{\rA\rA'} \otimes \M_{b'}^{\dag \rB\rB'} \M_b^{\rB\rB'} \otimes \mathds{1}^{\rB''}\right) \left({\phi^+}^{\rA''\rA'} \otimes {\varrho'}^{\rA\rB} \otimes {\phi^+}^{\rB'\rB''}\right) \right] (U_{a'}^{\dag \rA''} \otimes U_{b'}^{\dag \rB''}) \\ \nonumber
&=& \sum_{a,b,a',b'} \delta_{aa'} \delta_{bb'}(U_a^{\rA''} \otimes U_b^{\rB''}) \tr_{\rA'\rA\rB\rB'}\left[\left(\mathds{1}^{\rA''} \otimes \M_a^{\rA'\rA} \otimes M_b^{\rB\rB'} \otimes \mathds{1}^{\rB''}\right) \left({\phi^+}^{\rA''\rA'} \otimes {\varrho'}^{\rA\rB} \otimes {\phi^+}^{\rB'\rB''}\right) \right] (U_{a'}^{\dag \rA''} \otimes U_{b'}^{\dag \rB''}) \\ \nonumber
&=& \sum_{a,b,a',b'}(U_a^{\rA''} \otimes U_b^{\rB''}) \tr_{\rA'\rA\rB\rB'}\left[\left(\mathds{1}^{\rA''} \otimes \M_a^{\rA'\rA} \otimes \M_b^{\rB\rB'} \otimes \mathds{1}^{\rB''}\right) \left({\phi^+}^{\rA''\rA'} \otimes {\varrho'}^{\rA\rB} \otimes {\phi^+}^{\rB'\rB''}\right) \right] (U_{a}^{\dag \rA''} \otimes U_{b}^{\dag \rB''}) \\ \nonumber
&=& \frac{1}{d^2} \sum_{a,b} (U_a^{\rA''} \otimes U_b^{\rB''}) \tr_{\rA\rB}
\left[\left(\left({\M_a}^{\rA''\rA}\right)^{T_{\rA''}}\otimes \left({\M_b}^{\rB''\rB}\right)^{T_{\rB''}}\right) \left(\mathds{1}^{\rA''}\otimes {\varrho'}^{\rA\rB} \otimes \mathds{1}^{\rB''} \right)\right] (U_a^{\dag \rA''} \otimes U_b^{\dag \rB''})\\ \label{4sm}
&=&  \frac{1}{d^2}\sum_{a,b}  (U_a^{\rA''} \otimes U_b^{\rB''}) {\tilde{\M}_{ab}}^T (U_a^{\dag \rA''} \otimes U_b^{\dag \rB''}),
\end{eqnarray}
where $\tilde{\M}_{ab} $ are the effective measurements. To get the second equality we used the cyclic property of the trace. The orthonormality of the projection operators is used to obtain the third equality. The fifth equality is a consequence of the identity
\begin{equation}\label{ident}
\textrm{Tr}_{\rB}\left[\left(\M^{\rA\rB}\otimes \mathds{1}^\mathsf{C}\right)\left(\mathds{1}^\rA\otimes {\phi^+}^{\rB\mathsf{C}}\right)\right] = \frac{1}{d}\left(M^{\rA\mathsf{C}}\right)^{T_\mathsf{C}}
\end{equation}
and Eq. (\ref{4sm}) just uses the definition of the effective measurement.
Finally one can obtain
\begin{align} \label{5sm}
\textrm{Tr}_{\rA\rA'\rB\rB'}\Big(\Phi({\varrho'}^{\rA\rB})\Big) &= \frac{1}{d^2}\sum_{a,b} \frac{1}{d^2}\ketbra{\psi^{\rA\rB}}{\psi^{\rA\rB}}\\ \label{6sm}
&= d^4 \frac{1}{d^4}\ketbra{\psi^{\rA\rB}}{\psi^{\rA\rB}}
\end{align}
where the first equality follows directly from the constraint (\ref{2sm}), while the second follows from (\ref{2sm}) and takes into account that there are $d^2$ different values of $a$ and $d^2$ different values of $b$, which counts $d^4$ elements in the sum in (\ref{5sm}).
Thus, $\textrm{Tr}_{\rA\rA'\rB\rB'}\Big(\Phi({\varrho'}^{\rA\rB})\Big)$ is a pure, normalised state. We conclude that there is no entanglement between $\rA\rA'\rB\rB'$ and $\rA''\rB''$. Therefore, we can write
\[
\Phi( {\varrho'}^{\rA\rB}) =  \ketbra{\psi^{\rA''\rB''}}{\psi^{\rA''\rB''}} \otimes \varrho_{junk}^{\rA'\rA\rB\rB'}
\]

\section{Self-testing of multipartite states}\label{mltp}

The bipartite result is straightforwardly generalized to the multipartite case. Before stating the theorem let us define the scenario. There are $n$ parties (denoted by $\rA_1, \dots, \rA_n$) and they share the state ${\varrho'}^{\rA_1\dots \rA_n}$. Each of the parties can prepare a set of quantum inputs $\{\psi_{x_i}\}_i$, performs a joint measurement $\{\M_{a_i'}^{\rA_i}\}$ on the quantum input and its share of the state ${\varrho'}^{\rA_1\dots \rA_n}$ and returns the output $a_i$.\\
\textbf{Theorem 2} $\quad$
\textit{Let $n$ parties share the state ${\varrho'}^{\rA_1\dots \rA_n}$ and each of them has access to a tomographically complete set of inputs $\{{\psi_{x_i}}\}_{x_i}$ for $i = 1, \dots n$. Let the correlation probabilities obtained by measurements performed by each party have the form
\begin{equation}\label{7sm}
p(a_1,\dots, a_n\vert \psi_{x_1},\dots, \psi_{x_n}) =  \emph{Tr}\left[\tilde{\M}_{a_1,\dots, a_n}\left({\psi_{x_1}}\otimes \dots \otimes {\psi_{x_n}} \right)\right], 
\end{equation}
for every $x_i$. $\tilde{\M}_{a_1,\dots, a_n}$ are rank-one positive operators such that
\begin{equation}\label{8sm}
\tilde{\M} \equiv  (U_{a_1}\otimes \dots \otimes U_{a_n}){\tilde{\M}_{a_1,\dots, a_n}}^T(U_{a_1}^{\dagger}\otimes \dots \otimes U_{a_n}^{\dagger}) 
\end{equation}
for every $a_i$, where $U_{a_i}$ is the correcting unitaries generating the Bell state basis. Moreover, let
\begin{equation}\label{9sm}
\tilde{\M} = \frac{1}{d^{n}}\ketbra{\psi^{\rA_1\dots \rA_n}}{\psi^{\rA_1\dots \rA_n}}
\end{equation}
with $d^n$ being the dimension of the whole space of the $n$ systems. Then there exists a local isometry $\Phi$ such that
\begin{equation*}
\Phi({\varrho'}^{\rA_1\dots \rA_n}) = \ketbra{\psi^{\rA_1\dots \rA_n}}{\psi^{\rA_1\dots \rA_n}} \otimes\varrho_{junk}
\end{equation*}} \qed \\

\begin{proof}
The isometry used in the proof is just the multipartite generalization used in the bipartite case (see Fig. \ref{isomulti})

\begin{figure}
  \centerline{
    \begin{tikzpicture}[thick]
    %
    \tikzstyle{operator} = [draw,fill=white,minimum size=1.5em] 
    \tikzstyle{operator2} = [draw,fill=white,minimum height=1.8cm]
    \tikzstyle{phase} = [fill,shape=circle,minimum size=5pt,inner sep=0pt]
    \tikzstyle{circlewc} = [draw,minimum width=0.3cm]
    %
    \node at (0,0) (q1) {$\ket{0}^{\rA_i''}$};
    \node at (0,-1) (q2) {$\ket{0}^{\rA_i'}$};
    \node at (0,-2.5) (q3) {$\ket{\psi}^{\rA_1 \cdots \rA_n}$};
    \node at (0, -2) (qex1) {};
    \node at (3.23, -1) (qex3) {};
    \node at (3.25, -2) (qex4) {};
    \node at (2.80, -1) (qex5) {};
    \node at (2.80, -2) (qex6) {};
    %
    \node[operator] (op11) at (1,0) {F} edge [-] (q1);
    %
    \node[phase] (phase11) at (2,0) {} edge [-] (op11);
    \node[circlewc] (circlewc12) at (2,-1) {$\Sigma_x$} edge [-] (q2);
    \draw[-] (phase11) -- (circlewc12);
    \draw[-] (circlewc12) -- (qex5);
    \draw[-] (qex1)-- (qex6);
    %
    \node[operator] (op31) at (3,0) {$U_{a_i}$} edge [-] (phase11);
    \node[operator2] (operator32) at (3, -1.5) {$\M_{a_i}$}; 
[-] (circlewc42);
    \draw[-] (op31) -- (operator32);
    %
    \node (end1) at (5,0) {} edge [-] (op31);
    \node (end2) at (5,-1) {} edge [-] (qex3);
    \node (end3) at (5,-2) {} edge [-] (qex4);
    \end{tikzpicture}
  }
  \caption{
   Representation of one branch of the isometry $\Phi$. It takes as an input the state $\ket{\psi}^{\rA_1 \cdots \rA_n}$ and each party performs a unitary operation $U_{a_i}$ conditioned on the outcome of the measurement $M_{a_i}$. 
  }
  \label{isomulti}
\end{figure}
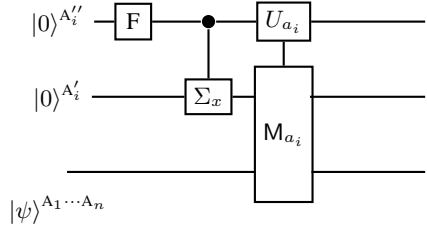
 Just as before
\begin{eqnarray*}
&\quad&\Phi({\varrho'}^{\rA_1 \dots \rA_n}) =  \sum_{a_i, {a_i}'} (U_{a_1}^{{\rA_1}''} \otimes \dots \otimes U_{a_n}^{{\rA_n}''} \otimes \M_{a_1}^{\rA_1{\rA_1}'} \otimes \cdots \otimes M_{a_n}^{\rA_n{\rA_n}'}) \\
&\quad& ({\Phi^+}^{{\rA_1}'{\rA_1}''} \otimes \dots \otimes {\Phi^+}^{{\rA_n}'{\rA_n}''}\otimes {\varrho'}^{{\rA_1} \dots {\rA_n}}) (U_{{a_i}'}^{\dag {\rA_1}''} \otimes \dots \otimes U_{{a_n}'}^{\dag {\rA_n}''} \otimes \M_{{a_1}'}^{\dag \rA_1{\rA_1}'} \otimes \dots \otimes \M_{{a_n}'}^{\dag \rA_n{\rA_n}'}).
\end{eqnarray*}
Tracing over $\rA_1,\rA_1', \cdots, \rA_n,\rA_n' $ we obtain
\begin{equation}\label{10sm}
\tr_{\rA_1,\rA_1', \dots, \rA_n,\rA_n'} \left(\Phi\left({\varrho'}^{\rA_1 \dots \rA_n}\right)\right) = \frac{1}{d^n} \sum_{a_i} (U_{a_1}^{\rA_1''} \otimes \dots \otimes U_{a_n}^{\rA_n''}) \tilde{\M}_{a_1,\dots, a_n}^T  (U_{a_1}^{\dag \rA_1''} \otimes \dots \otimes U_{a_n}^{\dag \rA_n''}),
\end{equation}
where $\tilde{\M}_{a1,\cdots, an}$ are the effective measurements satisfying constraints (\ref{7sm}-\ref{9sm}). Since every party has access to a tomographically complete set of quantum input states, (\ref{9sm}) is the only solution for $\tilde{M}$.\\ 
Finally,
\begin{eqnarray}\label{11sm}
\tr_{\rA_1,\rA_1', \dots, \rA_n,\rA_n'}\Phi({\varrho'}^{\rA_1, \dots,\rA_n}) &=& \sum_{a_1 ,\cdots, a_n} \frac{1}{d^n}  \tilde{\M} \\ \label{12sm}
&=& d^{2n}\frac{1}{d^n}\frac{1}{d^{n}}\ketbra{\psi^{\rA_1\dots \rA_n}}{\psi^{\rA_1\dots \rA_n}},
\end{eqnarray}
where the first equality comes from (\ref{10sm}) and (\ref{8sm}), while the second is a direct consequence of (\ref{9sm}). Since $\tr_{\rA_1,\rA_1', \dots, \rA_n,\rA_n'}\Phi({\varrho'}^{\rA_1, \dots,\rA_n})$ is a normalised pure state  $\Phi(\varrho^{\rA_1, \cdots,\rA_n})$ can be written as
\begin{equation*}
\Phi({\varrho'}^{\rA_1, \dots,\rA_n}) = \ketbra{\psi^{\rA_1\dots \rA_n}}{\psi^{\rA_1\dots \rA_n}}  \otimes \varrho_{junk}.
\end{equation*}

\end{proof}

%

\section{Self-testing of maximally entangled pair of qubits through CHSH inequality}\label{CHSH}

We already placed self-testing with quantum inputs in between quantum state tomography and standard self-testing. Self-testing with a tomographically complete set of quantum inputs was in spirit close to quantum state tomography (being based on the exact recovery of the effective measurement operators). One can ask if the approach closer to standard self-testing can be used for MDI recovery of quantum states. The idea is simple: if a set of projectors is used to produce measurement correlations that self-test a given state $\ket{\psi}$, they can be obtained by performing a Bell state measurement and preparing suitable inputs. For example
\begin{equation*}
\ketbra{0}{0}^A = \tr_{\rA'}\left[\left(\ketbra{0}{0}^{\rA'}\otimes \mathds{1}^\rA\right)({\phi^+}^{\rA'\rA} + {\phi^-}^{\rA'\rA})\right],
\end{equation*}
and 
\begin{equation*}
\ketbra{1}{1}^\rA = \tr_{\rA'}\left[\left(\ketbra{0}{0}^{\rA'}\otimes \mathds{1}^\rA\right)({\psi^+}^{\rA'\rA} + {\psi^-}^{\rA'\rA})\right].
\end{equation*}
Knowing this property the question is if the self-testing correlation probabilities obtained in a scenario with quantum inputs still self-test the state and moreover, if they self-test the Bell state measurement. Intuitively, the answer should be positive, since self-testing correlations can be obtained only if specific measurements are applied to a specific state. The formalization of this intuition for the case of the two-qubit maximally entangled state is given in the following theorem. For the sake of simplicity let us self-test the following form of the state
\begin{equation}
\ket{\psi} = \cos\left(\frac{\pi}{8}\right)\ket{\phi^+} +\sin\left(\frac{\pi}{8}\right)\ket{\psi^+},
\end{equation}
since it is locally unitarilly equivalent to $\ket{\phi^+}$, but both parties apply Pauli measurements to maximally violate CHSH inequality. \\
\textbf{Theorem 4} $\quad$
\textit{Let two parties, Alice and Bob, share the state ${\varrho'}^{\rA\rB}$ and let both Alice and Bob have access to the set of quantum inputs $\{\psi_0 = \ket{0}, \psi_1 = \ket{+}\}$. Alice and Bob apply a four-outcome measurement $\{M_a^{\rA'\rA}\}_a$ and $\{M_b^{\rB'\rB}\}_b$, respectively. Furthermore, let the following correlations hold
\begin{align} \label{chsh}
\bra{\psi'}A_0B_0 + A_0 &B_1  + A_1B_0 - A_1B_1\ket{\psi'} = 2\sqrt{2},
\end{align}
where 
\begin{subequations}\label{arya}
\begin{align}\nonumber
X_0 &= X_0^+ - X_0^-; \quad  X_1 = X_1^+ - X_1^-; \\ \nonumber
X_0^+ &= \textrm{tr}_{\rX'}\left[\left(\psi_0^{\rX'}\otimes \mathds{1}^\rX\right)\left(\M_0^{\rX'\rX}+\M_1^{\rX'\rX}\right)\right] \\ \nonumber
X_0^- &= \textrm{tr}_{\rX'}\left[\left(\psi_0^{\rX'}\otimes \mathds{1}^\rX\right)\left(\M_2^{\rX'\rX}+\M_3^{\rX'\rX}\right)\right] \\ \nonumber
X_1^+ &= \textrm{tr}_{\rX'}\left[\left(\psi_1^{\rX'}\otimes \mathds{1}^\rX\right)\left(\M_0^{\rX'\rX}+\M_2^{\rX'\rX}\right)\right] \\ \nonumber
X_1^- &= \textrm{tr}_{\rX'}\left[\left(\psi_1^{\rX'}\otimes \mathds{1}^\rX\right)\left(\M_1^{\rX'\rX}+\M_3^{\rX'\rX}\right)\right] 
\end{align}
\end{subequations}
where $X \in \{A,B\}$.
Then there is a local isometry $\Phi$ such that
\begin{eqnarray}
\Phi({\varrho'}^{\rA\rB}) = {\phi^+}\otimes \varrho_{\textrm{junk}}.
\end{eqnarray}
}

 This theorem represents the analogue of the self-testing of maximally entangled pair of qubits via the maximal violation of the CHSH (Clauser-Horn-Shimony-Holt) inequality. 
 \begin{proof}
 Let us first verify that $\{A_j^+,A_j^-\}$ for $j = 0,1$ represent valid measurements. Positivity is ensured by the relation
\begin{equation*}
\bra{\xi}{A_0^+}\ket{\xi} = \tr\left[\left(\psi_0^{\rA'}\otimes \xi^\rA\right)\left(\M_0^{\rA'\rA}+\M_1^{\rA'\rA}\right)\right] \geq 0, \qquad \forall \xi \geq 0,
\end{equation*} 
and similarly for the other operators $A_j^{\pm}$. The completeness relation is also satisfied:
\begin{eqnarray*}
A_j^+ + A_j^- &=& \tr_{\rA'}\left[\left(\psi_j^{\rA'}\otimes \mathds{1}^\rA\right)\left(\sum_k\M_k^{\rA'\rA}\right)\right] \\ 
&=& \tr_{\rA'}\left[\left(\psi_j^{\rA'}\otimes \mathds{1}^\rA\right)\mathds{1}^{\rA'\rA}\right] = \mathds{1}^{\rA}
\end{eqnarray*}
In an analogue way one can prove that $B_j$ are valid measurement observables.
This is basically enough to prove the self-testing theorem, since we have that the two parties use valid quantum measurements to maximally violate the CHSH inequality. This means that there must exist a local isometry mapping the state $\varrho'$ to the maximally entangled pair of qubits. 
\end{proof}

\section{Proof of Theorem \ref{statementqc}}\label{proof3}

Let us, for convenience, first restate the theorem here.

\textbf{Theorem 3} $\quad$\label{statementqSM}
\textit{Let two parties, Alice and Bob, share a  state ${\varrho'}^{\rA\rB}$. Furthermore, let Alice use quantum inputs $\psi_0 = \ketbra{0}{0}$, $\bar{\psi}_0 = \ketbra{1}{1}$, $\psi_1 = \ketbra{+}{+}$ and $\bar{\psi}_1 = \ketbra{-}{-}$. If they observe $\mathcal{I}_{qc} = 4$ where $\mathcal{I}_{qc}$ is defined as 
\begin{equation}\label{iqcSM}
\begin{split}
\mathcal{I}_{qc} =  \sum_{a=0,1}(p(a,0|\psi_0,0)+p(a,1|\bar{\psi}_0,0))+\sum_{a=2,3}(p(a,1|\psi_0,0)+p(a,0|\bar{\psi}_0,0) )+ \\
 + 
 \sum_{a=0,2}(p(a,0|\psi_1,1)+p(a,0|\bar{\psi}_1,1))+\sum_{a=1,3}(p(a,1|\psi_1,1)+p(a,0|\bar{\psi}_1,1) ).
 \end{split}
\end{equation}
then there exists a local isometry $\Phi$ such that 
\begin{equation}\label{isomeqcSM}
\Phi_{qc}({\varrho'}^{\rA\rB}) = \ketbra{\phi^+}{\phi^+}\otimes \varrho_{junk}
\end{equation}}
\qed

Note that the expression $\mathcal{I}_{qc}$ can be seen as as a sum of four terms, each of them being itself a sum to obtain few particular outcomes for certain quantum and classical inputs. The fact that everything sums up to $4$ means that each term must be equal to $1$, i.e.:
\begin{equation}\label{eqs1}\begin{split}
p(0,0|\psi_0,0) + p(1,0|\psi_0,0) + p(2,1|\psi_0,0) + p(3,1|\psi_0,0) = 1,\\
p(0,1|\bar{\psi}_0,0) + p(1,1|\bar{\psi}_0,0) + p(2,0|\bar{\psi}_0,0) + p(3,0|\bar{\psi}_0,0) = 1,\\
p(0,0|\psi_1,1) + p(2,0|\psi_1,1) + p(1,1|\psi_1,1) + p(3,1|\psi_1,1) = 1,\\
p(0,1|\bar{\psi}_1,1) + p(2,1|\bar{\psi}_1,1) + p(1,0|\bar{\psi}_1,1) + p(3,0|\bar{\psi}_1,1) = 1.\end{split}
\end{equation}
Since we fix $\ket{\psi_0} = \ket{0}$, $\ket{\bar{\psi}_0} = \ket{1}$, $\ket{\psi_1} = \ket{+}$ and $\ket{\bar{\psi}_1} = \ket{-}$ Eqs.~\eqref{eqs1} imply
\begin{equation}\label{eqs2}\begin{split}
    \tilde{\M}_{0,0|0} + \tilde{\M}_{1,0|0} + \tilde{\M}_{2,1|0} + \tilde{\M}_{3,1|0} = \ketbra{0}{0},\\
    \tilde{\M}_{2,0|0} + \tilde{\M}_{3,0|0} + \tilde{\M}_{0,1|0} + \tilde{\M}_{1,1|0} = \ketbra{1}{1},\\
    \tilde{\M}_{0,0|1} + \tilde{\M}_{2,0|1} + \tilde{\M}_{1,1|1} + \tilde{\M}_{3,1|1} = \ketbra{+}{+},\\
    \tilde{\M}_{1,0|1} + \tilde{\M}_{3,0|1} + \tilde{\M}_{0,1|1} + \tilde{\M}_{2,1|1} = \ketbra{-}{-},
\end{split}
\end{equation}
where 
\begin{equation*}
    \tilde{\M}_{a,b|y}^{\rA'} = \textrm{Tr}_{\rA\rB}\left[\left(\M_a^{\rA'\rA}\otimes \M_b^\rB\right)\left(\mathds{1}^{\rA'}\otimes\varrho^{\rA\rB}\right)\right].
\end{equation*}
Since all the operators $\tilde{\M}_{a,b|y}$ are positive, they all must be proportional to the corresponding projector ($\tilde{\M}_{0,0|0}$ to $\ketbra{0}{0}$,$\tilde{\M}_{0,1|0}$ to $\ketbra{1}{1}$ and so on). The no-signalling condition, expressed as
\begin{equation}\label{nsgl}
\sum_b\tilde{\M}_{a,b|0} = \sum_b\tilde{\M}_{a,b|1} ,
\end{equation}
for all values of $a$ imposes certain constraints on the traces of the effective measurements. Denote
\begin{eqnarray*}
\tilde{\M}_{a,0|0} = \mu_{a,0}\ketbra{0}{0}, &\quad& \textrm{for}\quad a = 0,1 \qquad \tilde{\M}_{a,0|0} = \mu_{a,0}\ketbra{1}{1}, \quad \textrm{for}\quad a = 2,3  \qquad \\ \tilde{\M}_{a,1|0} = \mu_{a,1}\ketbra{0}{0}  &\quad& \textrm{for}\quad a = 2,3 \qquad
\tilde{\M}_{a,1|0} = \mu_{a,1}\ketbra{1}{1}  \quad \textrm{for}\quad a = 0,1
\\
\tilde{\M}_{a,0|1} = \nu_{a,0}\ketbra{+}{+}, &\quad& \textrm{for}\quad a = 0,2 \qquad \tilde{\M}_{a,0|1} = \nu_{a,0}\ketbra{-}{-}, \quad \textrm{for}\quad a = 1,3  \qquad \\ \tilde{\M}_{a,1|1} = \nu_{a,1}\ketbra{+}{+}  &\quad& \textrm{for}\quad a = 1,3 \qquad
\tilde{\M}_{a,1|1} = \nu_{a,1}\ketbra{-}{-}  \quad \textrm{for}\quad a = 0,2 \end{eqnarray*}
The condition \eqref{nsgl} imposes $\nu_{a,0} = \nu_{a,1} = \mu_{a,0} = \mu_{a,1}$ for all $a$. By plugging this in Eqs. \eqref{eqs2}  we obtain:
\begin{equation}\label{cond}
    \mu_{0,0} + \mu_{1,0} + \mu_{2,0} + \mu_{3,0} = 1
\end{equation}
Another no-signalling condition%
\begin{equation}\label{nsgl1}
\sum_a\tilde{\M}_{a,b|0} = \sum_a\tilde{\M}_{a,b|1} ,
\end{equation}
gives
\begin{equation}
    (\mu_{0,0}+\mu_{1,0})\ketbra{0}{0} + (\mu_{2,0}+\mu_{3,0})\ketbra{1}{1} = (\mu_{0,0}+\mu_{2,0})\ketbra{+}{+} + (\mu_{1,0}+\mu_{3,0})\ketbra{-}{-}  
\end{equation}
which implies $\mu_{a,0} = 0.25$ for all $a$.

The above given conditions imply correctiness of the following expressions
\begin{equation}\begin{split}
    \textrm{Tr}_{\rA\rB}\left[\left((\M_0 + U_1\M_1U_1^{\da} + U_2\M_2U_2^\da + U_3\M_3U_3^\da)^{\rA'\rA}\otimes \M_{0|0}^\rB\right)\left(\frac{\mathds{1}}{2}^{\rA'}\otimes{\varrho'}^{\rA\rB}\right)\right] = \frac{1}{2}\ketbra{0}{0},\\
    \textrm{Tr}_{\rA\rB}\left[\left((\M_0 + U_1\M_1U_1^{\da} + U_2\M_2U_2^\da + U_3\M_3U_3^\da)^{\rA'\rA}\otimes \M_{1|0}^\rB\right)\left(\frac{\mathds{1}}{2}^{\rA'}\otimes{\varrho'}^{\rA\rB}\right)\right] = \frac{1}{2}\ketbra{1}{1},\\
     \textrm{Tr}_{\rA\rB}\left[\left((\M_0 + U_1\M_1U_1^{\da} + U_2\M_2U_2^\da + U_3\M_3U_3^\da)^{\rA'\rA}\otimes \M_{0|1}^\rB\right)\left(\frac{\mathds{1}}{2}^{\rA'}\otimes{\varrho'}^{\rA\rB}\right)\right] = \frac{1}{2}\ketbra{+}{+},\\
    \textrm{Tr}_{\rA\rB}\left[\left((\M_0 + U_1\M_1U_1^{\da} + U_2\M_2U_2^\da + U_3\M_3U_3^\da)^{\rA'\rA}\otimes \M_{1|1}^\rB\right)\left(\frac{\mathds{1}}{2}^{\rA'}\otimes{\varrho'}^{\rA\rB}\right)\right] = \frac{1}{2}\ketbra{-}{-},
    \end{split}
\end{equation}
where $U_1 = \sigma_z^{\rA'}\otimes\mathds{1}^{\rA}$, $U_2 = \sigma_x^{\rA'}\otimes\mathds{1}^{\rA}$ and $U_3 = (\sigma_z\sigma_x)^{\rA'}\otimes\mathds{1}^{\rA}$.
These, further, can be rewritten as
\begin{equation}\label{proj}
    \begin{split}
        \tr_{\rB}\left[\left(\mathds{1}^{\rA'}\otimes \M_{0|0}^\rB\right)\tilde{\varrho}^{\rA'\rB}\right] = \frac{1}{2}\ketbra{0}{0}, &\qquad \tr_{\rB}\left[\left(\mathds{1}^{\rA'}\otimes \M_{1|0}^\rB\right)\tilde{\varrho}^{\rA'\rB}\right] = \frac{1}{2}\ketbra{1}{1},\\
        \tr_{\rB}\left[\left(\mathds{1}^{\rA'}\otimes \M_{0|1}^\rB\right)\tilde{\varrho}^{\rA'\rB}\right] = \frac{1}{2}\ketbra{+}{+}, &\qquad \tr_{\rB}\left[\left(\mathds{1}^{\rA'}\otimes \M_{1|1}^\rB\right)\tilde{\varrho}^{\rA'\rB}\right] = \frac{1}{2}\ketbra{-}{-},
    \end{split}
\end{equation}
where
\begin{equation*}
    \tilde{\varrho}^{\rA'\rB} = \tr_{\rA}\left[\left((\M_0 + U_1\M_1U_1^{\da} + U_2\M_2U_2^\da + U_3\M_3U_3^\da)^{\rA'\rA}\otimes \mathds{1}^\rB\right)\left(\frac{\mathds{1}}{2}^{\rA'}\otimes{\varrho'}^{\rA\rB}\right)\right]
\end{equation*}
Eqs. \eqref{proj} imply
\begin{equation}\label{oper}
    \begin{split}
        \tr\left[(\sigma_\mathsf{z}^{\rA'}\otimes B_0)\tilde{\varrho}^{\rA'\rB} \right] = 1,\\
        \tr\left[(\sigma_\mathsf{x}^{\rA'}\otimes B_1)\tilde{\varrho}^{\rA'\rB} \right] = 1,
    \end{split}
\end{equation}
where $B_0 = \M_{0|0}-\M_{1|0}$ and $B_1 = \M_{0|1}-\M_{1|1}$.
The operators of type $\tr_{\rA}\left[\left( U_a^{A'}\M_a^{\rA'\rA}{U_a^{\da}}^{\rA'}\otimes \mathds{1}^\rB\right)\left(\frac{\mathds{1}}{2}^{\rA'}\otimes{\varrho'}^{\rA\rB}\right)\right]$ are not positive in general (see \cite{isc2}), but they have a positive expectation value on all separable vectors $\sum_i\pi_i^{\rA'}\otimes\tau_i^\rB$. Furthermore, 
\begin{equation*}
    \tr\tilde{\varrho}^{\rA'\rB} = \sum_a\mu_{a,0} = 1
\end{equation*}
Thus, Eqs. \eqref{oper} imply
\begin{equation}\label{zz}\begin{split}
\left(\sigma_\mathsf{z}^{\rA'}\otimes \mathds{1}^\rB\right)\tilde{\varrho}^{\rA'\rB} &= \left(\mathds{1}^{\rA'}\otimes B_0\right)\tilde{\varrho}^{\rA'\rB},\\
\left(\sigma_\mathsf{x}^{\rA'}\otimes \mathds{1}^\rB\right)\tilde{\varrho}^{\rA'\rB} &= \left(\mathds{1}^{\rA'}\otimes B_1\right)\tilde{\varrho}^{\rA'\rB}\end{split} .
\end{equation}
These equations allow one to conclude that $B_0$ and $B_1$ anticommute
\begin{equation}\label{ac}
    \{B_0,B_1\}\varrho_\rB = 0
\end{equation}
where $\varrho_\rB$ is the reduced state of ${\varrho'}^{\rA'\rB}$.
From equations
\begin{equation}
    \begin{split}
        \tr\left[\left(\mathbb{M_z}^{\rA'\rA}\otimes B_0\right)\left(\ketbra{0}{0}^{\rA'}\otimes{\varrho'}^{\rA\rB}\right)\right] = 1,\\
        \tr\left[\left(\mathbb{M_x}^{\rA'\rA}\otimes B_1\right)\left(\ketbra{+}{+}^{\rA'}\otimes{\varrho'}^{\rA\rB}\right)\right] = 1
    \end{split} ,
\end{equation}
we can conclude that
\begin{equation}\label{aa}
    \begin{split}
        \left(\tr_{\rA'}\left((\ketbra{0}{0}^{\rA'}\otimes\mathds{1}^A)\mathbb{M_z}^{\rA'\rA}\right)\otimes \mathds{1}^\rB\right){\varrho'}^{\rA\rB} = \left(\mathds{1}^\rA\otimes B_0\right){\varrho'}^{\rA\rB},\\
        \left(\tr_{\rA'}\left((\ketbra{+}{+}^{\rA'}\otimes\mathds{1}^\rA)\mathbb{M_x}^{\rA'\rA}\right)\otimes \mathds{1}^\rB\right){\varrho'}^{\rA\rB} = \left(\mathds{1}^\rA\otimes B_1\right){\varrho'}^{\rA\rB} .
    \end{split}
\end{equation}
Finally, Eqs. \eqref{zz}, \eqref{ac}, \eqref{aa} allow reducing the expression $\tr_{\rA'_1,\rA'_2}\left[\Phi_{qc}({\varrho'}^{\rA\rB})\right]$, where $\Phi_{qc}$ is the circuit given in Fig. \ref{isoQCSM}, to the output of the standard self-testing SWAP gate, giving
\begin{equation}
    \tr_{\rA'_1,\rA'_2}\left[\Phi_{qc}({\varrho'}^{\rA\rB})\right] = \phi_+^{\rA''\rB'}\otimes \varrho_{junk}^{\rA\rB} .
\end{equation}
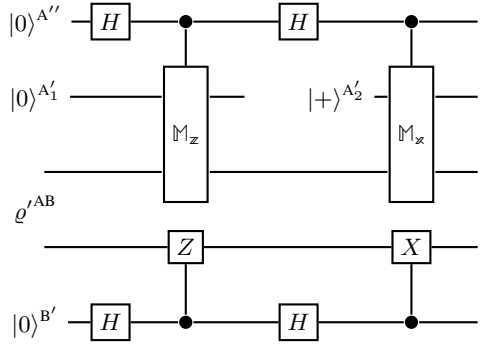
\begin{figure}
  \centerline{
    \begin{tikzpicture}[thick]
    %
    \tikzstyle{operator} = [draw,fill=white,minimum size=1.5em] 
    \tikzstyle{operator2} = [draw,fill=white,minimum height=1.8cm]
    \tikzstyle{phase} = [fill,shape=circle,minimum size=5pt,inner sep=0pt]
    \tikzstyle{circlewc} = [draw,minimum width=0.3cm]
    %
    \node at (0,0) (q1) {$\ket{0}^{\rA''}$};
    \node at (0,-1) (q2) {$\ket{0}^{\rA_1'}$};
    \node at (0,-2.5) (q3) {${\varrho'}^{\rA\rB}$};
    \node at (0, -2) (qex1) {};
    \node at (0, -3) (qex2) {};
    \node at (0,-4) (q4) {$\ket{0}^{\rB'}$};
    \node at (1.8, -1) (qex3) {};
    \node at (1.8, -2) (qex4) {};
    \node at (2.2,-1) (qq1) {};
    \node at (2.9,-1) (qq2) {};
    \node at (4,-1) (qq3) {$\ket{+}^{\rA_2'}$};
    \node at (4.8,-1) (qq4) {};
    \node at (2.2,-2) (qq5) {};
    \node at (4.8,-2) (qq6) {};
    \node at (5.2, -1) (qq7) {};
    \node at (5.2, -2) (qq8) {};
    %
    \node[operator] (op11) at (1,0) {$H$} edge [-] (q1);
    \node[operator] (op41) at (1,-4) {$H$} edge [-] (q4);
    %
    \node[phase] (phase11) at (2,0) {} edge [-] (op11);
    \node[operator2] (circlewc12) at (2,-1.5) {$\mathbb{M_z}$};
    \draw[-] (phase11) -- (circlewc12);
    \draw[-] (qex3) -- (q2);
    \draw[-] (qex4) -- (qex1);
    \node[phase] (phase42) at (2,-4) {} edge [-] (op41);
    \node[circlewc] (circlewc13) at (2,-3) {$Z$} edge [-] (qex2);
    \draw[-] (phase42) -- (circlewc13);
    \draw[-] (qq2) -- (qq1);
    \node[operator] (op12) at (3.5,0) {$H$} edge [-] (phase11);
    \node[operator] (op42) at (3.5,-4) {$H$} edge [-] (phase42);
    \node[phase] (phase12) at (5,0) {} edge [-] (op12);
    \node[operator2] (circlewc22) at (5,-1.5) {$\mathbb{M_x}$};
    \draw[-] (phase12) -- (circlewc22);
    \draw[-] (qq4) -- (qq3);
    \draw[-] (qq5) -- (qq6);
    \node[phase] (phase42) at (5,-4) {} edge [-] (op42);
    \node[circlewc] (circlewc23) at (5,-3) {$X$} edge [-] (circlewc13);
    \draw[-] (phase42) -- (circlewc23);
    \node (end1) at (6,0) {} edge [-] (phase12);
    \node (end2) at (6,-1) {} edge [-] (qq7);
    \node (end3) at (6,-2) {} edge [-] (qq8);
    \node (end4) at (6,-3) {} edge [-] (circlewc23);
    \node (end5) at (6,-4) {} edge [-] (phase42);
    \end{tikzpicture}
  }
  \caption{
    Isometry $\Phi_{qc}$ used in the Proof of Theorem 3. 
  }
  \label{isoQCSM}
\end{figure}

\end{appendix}

\end{document}